\documentclass[11pt]{article}
\usepackage{geometry}
\usepackage{graphicx}
\usepackage{subfigure}
\usepackage{subcaption}
\usepackage{float}
\usepackage{multirow}
\usepackage{hyperref}
\usepackage{cite}
\newcommand{\cbl}{}

\begin{document}

\title{Numerical model for pellet rocket acceleration in PELOTON \footnote{Submitted to Plasma Physics and Controlled Fusion}}

\author{J. Corbett$^1$, R. Samulyak$^1$\footnote{roman.samulyak@stonybrook.edu}, F.J. Artola$^{2}$, S. Jachmich$^2$, M. Kong$^{3}$, E. Nardon$^{4}$\\
{$^1$Department of Applied Math. and Statistics, Stony Brook University, Stony Brook NY 11794 USA} \\
{$^2$ITER Organization, 13067 Saint Paul-lez-Durance, Cedex, France} \\
{$^3$École Polytechnique Fédérale de Lausanne (EPFL), CH-1015 Lausanne, Switzerland} \\
{$^4$CEA, IRFM, F-13108 Saint-Paul-lez-Durance, France } }
\date{January 2026}

\maketitle

\begin{abstract}
A direct numerical simulation model for the rocket acceleration of pellets in thermonuclear fusion devices has been developed for PELOTON, a 3D Lagrangian particle pellet code [R. Samulyak et al, Nuclear Fusion 61 (4), 046007 (2021)], and validated using shattered pellet injection (SPI) experiments in JET. The pellet rocket acceleration is driven by grad-$\mathbf{B}$ drift of the ablation cloud that creates asymmetry and non-uniform heating of the cloud. The model accounts for non-uniform charging of the ablation cloud by hot plasma electrons as well as local plasma gradients. As a result, the increased pressure on the high-field-side compared to the low-field-side leads to pellet (fragment) rocket acceleration. Pure deuterium and deuterium-neon mixture models have been implemented. The background plasma states have been obtained by using a new plasma cooling model for PELOTON. The cooling model distributes the ablated material within the corresponding flux volumes and accounts for ionization and other energy losses, Ohmic heating by toroidal currents, and the energy exchange between ions and electrons. Plasma profiles predicted by PELOTON's cooling model have been compared with JOREK and INDEX simulations. PELOTON simulations of rocket acceleration and the corresponding trajectories of deuterium fragments are consistent with experimentally measured trajectories in JET. We show that composite deuterium-neon pellets containing 0.5\% of neon experienced smaller deviation of their trajectories compared to the pure deuterium case. We simulate various spatial configurations of pellet fragments and demonstrate the cloud overlap impact on rocket acceleration. Additionally, we demonstrate the effect of plasma state gradients on the rocket acceleration. Future work will focus on the rocket acceleration of SPI in projected ITER plasmas and the development of the corresponding scaling law for the rocket acceleration. 
\end{abstract}


\section{Introduction}

\begin{figure}
	\centering
	\includegraphics[width=0.9\linewidth]{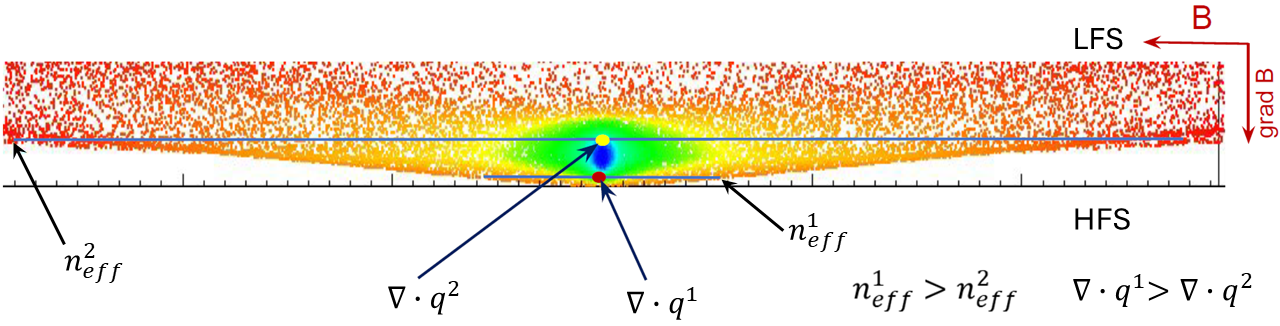}
	\caption{Schematic of non-uniform pellet cloud heating causing the rocket acceleration. $n_{eff}^1$ ($n_{eff}^2$) is the effective electron density on HFS (LFS) due to electrostatic shielding which reduces the intensity of electrons streaming into the ablation cloud along magnetic field lines. $\nabla\cdot q^1$ and $\nabla\cdot q^2$ denote the hot electron energy deposition in the cloud on HFS and LFS, correspondingly. See Section 2.2 for exact expressions.}
	\label{fig1}
\end{figure}

Rocket acceleration of cryogenic pellets or shattered pellet injection (SPI) fragments injected into a thermonuclear fusion device is a consequence of asymmetric shape of ablation clouds due to transport of the ablated material along and across magnetic field lines and their non-uniform heating by hot plasma electrons. The effect was observed experimentally starting from early experiments operating with hydrogen pellets \cite{JorgensenSillesen1975}.
As a cryogenic pellet is injected into plasma, it experiences an intense heating by hot background plasma electrons, streaming along magnetic field lines, and a rapid ablation. A cold, neutral, high-pressure cloud is formed around the pellet that initially expands spherically by pressure gradients and moves together with the pellet. The cloud then ionizes, flows along magnetic field lines in the pellet reference frame by the action of Lorentz force, and drifts across magnetic field lines by the magnetic field curvature-induced polarization $\mathbf{E}\times \mathbf{B}$ drift, called hereafter as grad-$\mathbf{B}$ drift, where $\mathbf{E}$ and $\mathbf{B}$ refer to the electric and magnetic field, respectively. The grad-$\mathbf{B}$ drift establishes a finite shielding length of the pellet cloud (defined here as the length of ablated cloud along the magnetic field line passing through the pellet center). In the absence of background plasma perturbations, the ablation reaches quasi-steady state (the time for establishing a steady state is shorter compared to a noticeable pellet radius reduction due to ablation). In this paper, we do not consider mechanisms for the periodic separation of ablated clouds from pellets, causing so-called striation instabilities (\cite{LangConway2024} and references therein), and assume that steady state ablation remains as long as the background plasma states and the pellet radius are constant. The grad-$\mathbf{B}$ drift bends the ablation cloud in the pellet vicinity (see Figure \ref{fig1}). Portions of the ablation cloud on the high-field-side (HFS) of the pellet provide less electrostatic shielding effect to the  incoming plasma electrons compared to the low-field-side (LFS), and the intensity of incoming plasma electrons is higher on the HFS. In addition, the incoming plasma electrons experience less decay due to shorter paths to any point in the ablation cloud on the HFS compared to the LFS (Figure \ref{fig1}). These two factors lead to increased pressure on the HFS of the pellet surface compared to the LFS and cause acceleration of pellet in the LFS direction (hereafter called as rocket acceleration):
\begin{equation}
	a=\frac{\left(P_{\rm{HFS}}-P_{\rm{LFS}}\right) \cdot r_p^2 \pi}{m_p},
\label{eq:acceleration}
\end{equation}
where $P$ is pressure near the pellet surface, $r_p$ is the pellet radius, and $m_p$ is the pellet mass. In addition, the rocket acceleration may be influenced by local plasma gradients which may increase or decrease $a$. 
The purpose of this work is to report a direct numerical simulation model for pellet rocket acceleration and its implementation in PELOTON, a Lagrangian particle-based code for the simulation of pellets and SPI fragments \cite{SamulyakYuan2021}, and perform numerical studies of the rocket acceleration. 

Pellet rocket acceleration has been examined in several theoretical studies. A semi-empirical rocket acceleration model was incorporated into the neutral gas shielding (NGS) ablation model and compared with experimental data from ASDEX-Upgrade \cite{SzepesiKalvin2009}. In that model, the pressure difference driving the rocket acceleration is treated as an adjustable parameter, which limits its predictive capability.

A semi-analytical model estimating the rocket force from the asymmetry of the ablation rate was developed in \cite{SenichenkovRozhansky2007}. A more advanced semi-analytical approach was recently presented in \cite{GuthVallhagen2025} (with a detailed version available on arXiv \cite{GuthVallhagenArXiv}), where perturbation series expansions for the density, pressure, and ablated material velocity around an ablating pellet were derived relative to the “ground state” given by NGS solutions. This model attempts to capture the effects of asymmetric pellet heating caused by grad-$\mathbf{B}$ drift–induced variations in cloud shielding length and local plasma gradients. While this model predicts radial distributions of ablated gas states around the pellet, they are not very realistic due to model approximations. The model focuses exclusively on the rocket effect and omits higher order terms in the perturbation series such as larger deviations of ablated gas states in the direction of magnetic field line passing through the pellet center (i.e. their first-order correction should properly appear at second order in a more comprehensive theory). For this reason, our simulations disagree with the distribution of states shown in  Figure 5 in \cite{GuthVallhagenArXiv}. But in terms of the rocket force, our simulations are in qualitative agreement with this model. In particular, both  \cite{GuthVallhagenArXiv} and our simulations predict that the rocket acceleration is mainly due to the pressure difference between the HFS and the LFS of the pellet. 
The momentum flux term due to ablation rate difference $\int\int_S \rho \mathbf{v}\mathbf{v}\cdot \mathbf{r} dS$, there $\mathbf{r}$ is the grad-$\mathbf{B}$ drift direction, $\mathbf{v}$ is the velocity vector at the pellet surface, and the integration is performed over the pellet surface, is found to be negligibly small in the PELOTON simulations. We explain this by the large gas density around the pellet, which strongly shields it from the asymmetric background plasma electron heat flux. The vast majority of the latter is converted into gas pressure. We also write (\ref{eq:acceleration}) in a simplified form rather than in the integral form because the pressure difference and its scaling with background plasma states (and not the acceleration itself) represent our main results usable in other codes.  Therefore, the term rocket acceleration does not properly represent the nature of this phenomenon. We continue using it but note that the pellet motion in the LFS direction resembles more a projectile driven by pressure behind it than a rocket that is mainly driven by the momentum flux of ejected gases.

The paper is structured as follows. Section 2 provides a detailed description of the systems of equations governing the pellet ablation process. The numerical implementation of the rocket acceleration model in PELOTON is briefly summarized in Section 3; readers are referred to \cite{SamulyakYuan2021} for a more detailed discussion of PELOTON’s numerical approximations. Section 4 presents and discusses the numerical results, including a scaling law derived from fitting numerical data of SPI simulations in JET, the rocket acceleration experienced by SPI fragments, their trajectories, and comparisons with experimental observations. The effects of the neon component in pellets, local plasma gradients, and partial cloud overlap are also quantified. Finally, Section 5 concludes the paper with a summary of the main findings and an outline of future work.

\section{Models and governing equations}

\subsection{Pellet surface ablation and low magnetic Reynolds number MHD approximation for ablation cloud}

The pellet surface ablation model, as in \cite{SamulyakYuan2021,SamulyakLu2007}, assumes that all electron energy that reaches the pellet surface is completely used for vaporization of the pellet material. Ablation on the pellet surface satisfies three boundary conditions. First, the heat diffusion in the solid pellet is slow compared to the pellet radius recession speed, therefore the pellet surface temperature remains constant. Second, the normal velocity of the ablated material at the surface $u$ is determined by the heat flux into the pellet, $q$, and the sublimation energy $\epsilon_{\mathrm{s}}$
\begin{equation}
	\frac{q}{\epsilon_{\mathrm{s}}}=\rho_{\mathrm{v}} u,
\end{equation}
where $\rho_{\mathrm{v}}$ is the vapor density at the pellet surface. The third condition is the characteristic hydrodynamic equation \cite{SamulyakWang2018} along the normal direction from the ablation cloud onto the pellet surface:
\begin{equation}
	\frac{\partial P}{\partial t}-c_s \frac{\partial P}{\partial n}-\rho c_s\left(\frac{\partial u}{\partial t}-c_s \frac{\partial u}{\partial n}\right)=(\gamma-1) \frac{\partial q_\pm}{\partial z}
\end{equation}
where $c_s$ is the sound speed in the cloud, $\gamma$ is the ratio of specific heats, $z$ is the direction of the electron flux, and $ q_\pm$ is the heat deposition on the pellet surface in the right (+) or left (-) direction described in Section \ref{sec:e_flux}.

Evolution of a cold, partially ionized ablation cloud is described by the system of MHD equations in the low Magnetic Reynolds number approximation, ${\delta B} / B \sim R_m << 1$, where $\delta B$ is the eddy current induced magnetic field. In Lagrangian coordinates, the equations are
\begin{eqnarray}
	\frac{d \rho}{d t}  = - \rho \nabla \cdot \mathbf{u}, \label{eq:mass}\\
	\rho \frac{d \mathbf{u}}{dt}= -\nabla P + \mathbf{J \times B} + \mathbf{f}_D, \label{eq:momentum}\\
	\rho \frac{d\epsilon}{dt} = -P\nabla \cdot \mathbf{u} + \frac{1}{\sigma_{\perp}}\mathbf{J}^2 -\nabla \cdot q,\label{eq:energy}\\
	P = P(\rho,\epsilon), \label{eq:eos}
\end{eqnarray}
where $d/dt$ is the Lagrangian time derivative, $\mathbf{u}$, $\rho$ and $\epsilon$ are the velocity, density and specific internal energy, respectively, $P$ is the pressure, $\mathbf{B} = (0,0,B_{||})$ is the prescribed magnetic field induction, $\mathbf{J}$ is the current density computed as $\mathbf{J} = \sigma_{\perp} \mathbf{u}\times \mathbf{B}$, and $\sigma_{\perp}$ is the transverse electric conductivity. Last term in (\ref{eq:energy}), the energy deposition of hot plasma electrons streaming into the ablation cloud, is described in Section \ref{sec:e_flux}. Finally, the ablation cloud experiences grad-$\mathbf{B}$ drift in the transverse direction to  the magnetic field {\cbl governed by force} $\mathbf{f}_D$ in (\ref{eq:momentum}). In present work, we use a simplified grad-$\mathbf{B}$ drift model as we are interested in the state of ablation cloud sufficiently close to the pellet in transverse direction to the magnetic field. 
In particular, the transverse drift acceleration $d v_{\mathrm{D}}/dt = \mathbf{f}_D/\rho$ in the LFS direction is governed by the following equation 
\begin{equation}
	\frac{\mathrm{d} v_{\mathrm{D}}}{\mathrm{~d} t}=\frac{2\left\langle P\left(1+\frac{M_{\|}^2}{2}\right)-P_{\infty}\right\rangle}{R\,\langle\rho\rangle}
	\label{eq:gradB}
\end{equation}
Here $\langle A\rangle$ is the integral of quantity $A$ along a magnetic field line, $\langle A\rangle \equiv \int_{-L}^L A \mathrm{~d} z$, where $L$ is the cloud length, $R$ denotes the magnetic field radius of curvature, $P_{\infty}$ is the ambient plasma pressure, and $\left\langle P-P_{\infty}\right\rangle$ is called the drive integral. 
We use a more advanced model for cross-field transport of \cite{YuanNaitlho2022} if modeling of the plasmoid drift over larger distances is necessary. {\cbl The Alfven wave drag and the external connection current drag terms considered in \cite{YuanNaitlho2022} are important at larger distances from the pellet in the transverse direction compared to length scales for computing ablation rates and rocket acceleration values.}

\subsection{Electron heat deposition}
\label{sec:e_flux}

The electron heat flux model is based on \cite{IshizakiParks2004,ParksBosviel2020}, with several improvements \cite{ZhangParks2020} relevant to high-Z elements, but it includes a more advanced model for the electrostatic shielding of the cloud / albedo effect from \cite{ParksLu2009} that is critical for pellet rocket acceleration.

Based on solutions of the 3-D linearized Fokker-Planck kinetic equation \cite{IshizakiParks2004}, the heat source $-\nabla \cdot q$ coming from the energy deposition by hot, long mean-free path electrons streaming into the ablation cloud along the magnetic field lines is
\begin{equation}
	-\nabla \cdot \mathbf{q}=\frac{q_{\infty} n_{\rm e}(r, z)}{\tau_{\rm eff}} \left[g\left(w_{+}\right)+g\left(w_{-}\right)\right],
\end{equation}
where
\begin{equation}
	q_{\infty}=\sqrt{\frac{2}{\pi m_{\rm e}}} n_{\rm eff}\left(k T_{{\rm e} \infty}\right)^{\frac32},
\end{equation}
and $n_{\rm eff}$ is the effective plasma electron density due to the electrostatic shielding/albedo effect,
\begin{equation}
	n_{\rm eff}=\left(1-\frac{A}{100}\right) e^{-\Phi} n_{\rm e\infty}.
\end{equation}
Here $e^{-\Phi}$ is the decrease due to electrostatic shielding, and the shielding potential $\Phi$ is described at the end of this section.
Quantity $A$ is the surface reflectivity due to collisional backscattering,
\begin{equation}
	A=23.92 \ln \left(1+0.2014\left(1+Z_*\right)\right),
\end{equation}
$g(w)=w^{\frac{1}{2}} K_1\left(w^{\frac{1}{2}}\right) / 4$, where $K_1$ is the standard modified Bessel function of the second kind and $T_{\mathrm{e} \infty}$ is the temperature of the plasma electrons. Quantities $w_{ \pm}=\frac{\tau_{ \pm}}{\tau_{\mathrm{cff}}}$ are dimensionless opacities of the cloud with respect to each (+ or -) longitudinal direction from a point of interest, where the respective line integrated densities of the ablation electrons (bound and free) are,
\begin{equation}
	\tau_{+}(r, x)=\int_{-\infty}^x n_{\mathrm{e}}\left(r, x^{\prime}\right) \mathrm{d} x^{\prime} \quad {\rm and} \quad  \tau_{-}(r, x)=\int_x^{\infty} n_{\mathrm{e}}\left(r, x^{\prime}\right) \mathrm{d} x^{\prime}.
\end{equation}

Hot electron energy flux is attenuated by a combination of slowing down and pitch angle scattering with an effective energy flux attenuation thickness given by
\begin{equation}
	\tau_{\mathrm{eff}}=\tau_{\infty} \frac{1}{0.625+0.55 \sqrt{1+Z_*}}, \quad \tau_{\infty}=\frac{T_{\mathrm{e} \infty}^2}{8 \pi e^4 \ln \Lambda},
\end{equation}
where $e$ is the elementary charge of electron {\cbl and $Z_*$ denotes the average ionization state of partially ionized plasma}. The Coulomb logarithm $\ln \Lambda$ here pertains to inelastic scattering of fast electrons off atomic (bound) electrons in the neutral gas target,
\begin{equation}
	\ln \Lambda=\ln \left(\frac{E}{I^*} \sqrt{\frac{e(=\exp(1))}{2}}\right),
\end{equation}
where $I^*$ is the mean excitation energy for neutral atoms. The Coulomb logarithm is evaluated at energy $E \approx 2 T_{\mathrm{e} \infty}$ since that is the average energy per particle in the cloud from the distribution of semi-isotropic incident Maxwellian electrons.

The electrostatic shielding potential $\Phi$, varying in a plane normal to the magnetic field and constant along each magnetic field line, is a solution of the following nonlinear equation \cite{ParksLu2009}
\begin{equation}
	\exp \left(-\Phi\right)-\sqrt{\pi \Phi} \, {\rm erfc}\left(\sqrt{\pi \Phi}\right)=\frac{\alpha}{1-\mu},
	\label{eq:phi}
\end{equation}
{\cbl where $\rm erfc$ is the Gauss error function} and the attenuation coefficient $\mu$ is a function of the normalized cloud opacity $w = w_{+} + w_{-}$
\begin{equation}
	\mu(w)=\left[1+ \sqrt{\frac w2} - \frac w4 + \frac12 \left(\frac w2\right)^{\frac32}\right] \exp\left(-\sqrt{\frac w2}\right)-\left(\frac{w^2}8\right) E_1(\sqrt{\frac w2}),
	\label{eq:mu}
\end{equation}
{\cbl where $E_1$ is the exponential integral}.

Finally, the heat deposition on the surface of the pellet is given by
\begin{equation}
	q_{\pm}=q_{\infty} \frac{1}{2} w_{ \pm} K_2\left(w_{ \pm}^{\frac{1}{2}}\right),
\end{equation}
where $q_{\pm}$ is either $q_{+}$ or $q_{-}$, the electron heat flux impacting the left or right sides of the pellet, correspondingly.

\subsection{Equation of state for deuterium-neon mixture}

Let $C$ denote the ratio of number of deuterium to neon nuclei. We wish to solve for the particle {\cbl fractions $f^0_{Ne}, f^1_{Ne}, \ldots, f^{10}_{Ne}, f_D, f_{D+}$, and $f_e$, where $f^i_{Ne}$ is the fraction of the i-th ionized Neon ion, $f_D$ is the fraction of deuterium atoms, $f_{D^{+}}$ is the fraction of deuterium ions, and $f_e$ is the fraction} of electrons. These fractions are used to compute quantities such as temperature, sound speed, and electric conductivity. We setup a nonlinear system of equations based on 10 Saha equations for neon, 2 Saha equations for deuterium, along with conservation of mass and conservation of charge equations. These equations are as follows:
\begin{eqnarray}
	\frac{f^{m+1}_{Ne} f_e}{f^m_{Ne}}=\frac{2}{N_t} \frac{g_{m+1}^N}{g_m^N}\left(\frac{2 \pi m_e k T}{h^2}\right)^{\frac{3}{2}} \exp \left(-\frac{I_{m+1}^N}{k T}\right), \quad m=0,1,\ldots, 9 \\
	\frac{f_{D^{+}} f_e}{1-f_{D^{+}}}=3 \times 10^{21} \frac{T^{\beta_i}}{N_t} \exp \left(-\frac{\epsilon_i}{T}\right), \\
	\frac{\left(f_D+f_{D^{+}}\right) f_e}{1-\left(f_D+f_{D^{+}}\right)}=1.55 \times 10^{24} \frac{T^{\beta_d}}{N_t} \exp \left(-\frac{\epsilon_d}{T}\right), \\
	f^0_{Ne}+f^1_{Ne}+f^2_{Ne}+\ldots+f^{10}_{Ne}=1 \\
	f_e=\frac{f^1_{Ne}}{1+C} + \frac{2 f^2_{Ne}}{1+C} + \ldots +
	\frac{10 f^{10}_{Ne}}{1+C} + \frac{f_{D^{+}}}{1+1/C},
\end{eqnarray}
where $h$ is the Planck constant, $k$ is the Boltzmann constant, $m_e$ is the electron mass, $g_m$ are known electron partition functions, and $I_m$ are the successive ionization potentials. For deuterium, the ionization energy is $\epsilon_i=13.6 \mathrm{eV}$ and the dissociation energy is $\epsilon_d=4.48 \mathrm{eV}$, and finally we have $\beta_i=\frac{3}{2}$ and the parameter $\beta_d=0.327$. 

This is a fully coupled system of nonlinear equations that is difficult to solve in each point at every timestep of our code. We developed a multispecies Saha equation solver designed to compute the necessary thermodynamic quantities in tabular form. For each pair of pressure and density $(P, \rho)$, the main steps of our multi-species Saha equation solver can be described as follow:
\begin{enumerate}
	\item We start with $P$ and $\rho$ and the solution of the nonlinear system is denoted $x=$ $\left(f_0^N, f_1^N, \ldots, f_{10}^N, f_D, f_{D+}, f_e\right)$. We first compute $N_t=\rho / m_{\rm avg}$, where $m_{\rm avg}=m_{N e o n}\left(\frac{1}{1+C}\right)+$ $m_{\rm Deuterium}\left(\frac{C}{1+C}\right), m_{\rm Neon}=3.351 \times 10^{-23} g$ and $m_{\rm Deuterium}=3.34 \times 10^{-24} g$.
	
	\item We make an initial guess $x_0$, which is chosen as the solution of the nonlinear system obtained from the previous pair of pressure and density in our table. This initial guess is used to find the particle temperature $T$ by
	\begin{equation}
		T=\frac{P}{\left(1+f_e\right) \rho {\cal R}}
	\end{equation}
	where ${\cal R}=83.14 / \mu_{\rm avg}$, $\mu_{\rm avg}=\mu_{\rm Neon}\left(\frac{1}{1+C}\right)+\mu_{\rm Deuterium}\left(\frac{C}{1+C}\right)$, 
	$\mu_{\rm Neon}=20.1797$ and $\mu_{\rm Deuterium}=2.014$. For the first point in the table, we compute $T$ using the ideal gas equation of state.
	
	\item Using the computed values of $N_t$ and $T$, we solve the system of equations for the particle concentrations in the vector $x$ using an iterative nonlinear system solver. The iterative process will be done on the convergence of these particle concentrations, where we iterate and solve the nonlinear system until $\left|x_{\rm new}-x_{\rm old}\right|$ is less than a prescribed tolerance. Note that the temperature also changes at each iteration and it converges as $x$ converges.
\end{enumerate}

The radiation power density $P_{\rm rad}$ for the deuterium-neon mixture is computed as
\begin{equation}
	P_{r a d}=n_e n_{N e} L.
\end{equation}
Here $n_{Ne}$ is the number density of neon nuclei and $n_e$ is the number density of free electrons computed as $n_e = n_D*f_{D,i} + Z\, n_{Ne}$,
where $n_D$ is the number density of deuterium nuclei and $f_{D,i}$ is the deuterium ionization rate. The average neon ionization state $Z$ and the effective radiation coefficient $L$ are also given in tabulated form as functions of only temperature, computed by the CRETIN code \cite{Scott2001}.

\subsection{Plasma Cooling Module}

PELOTON Cooling Module was developed \cite{CorbettSamulyak2025} to provide PELOTON with approximate evolving background plasma states during SPI for use in rocket acceleration studies. The module is implemented as a stand-alone code and can also be used for the ablated material deposition studies. 

PELOTON's Cooling Module implements a physics model that discretizes the plasma into flux volumes, distributes ablated material within the corresponding flux volumes and accounts for ionization and other energy losses including radiation, Ohmic heating by fixed  toroidal currents, and the energy exchange between ions and electrons. In simulation results performed by the cooling module presented herein, radiation modeling is not applied. Plasma profiles predicted by PELOTON’s cooling model have been compared with simulations from two other plasma evolution codes, JOREK and INDEX \cite{MatsuyamaHu2022,KongNardon2024}. {\cbl PELOTON's Cooling Module differs from INDEX by using a temporally fixed flux surface geometry and current profile. Additionally, transport and radiation models are not applied in this study. PELOTON's Cooling Module incorporates pellet ablation and  rocket acceleration models referencing data from PELOTON simulations.}

PELOTON's Cooling Module represents the plasma using a number of discrete flux volumes, each of which has a spatially constant thermodynamic state. Consider an ablating SPI fragment moving with a certain velocity in a tokamak. During an interval of time $d t$, chosen sufficiently small so that the background plasma temperature and density can be considered constant during the interval, the pellet ablates material that increases the ion and electron densities by $dn_i$ and $dn_e$, respectively. These quantities are computed {\cbl by solving the following equations in the time interval $d t$}:
\begin{equation}
	{\cbl \frac{dn_i}{dt} = \frac{G}{V\,m_i}, \quad \frac{dn_e}{dt} = f_e \frac{dn_i}{dt},}
	\label{eq:material deposition}
\end{equation} 
where $G$ is the SPI fragment ablation rate (in kg/s) computed by (or with reference to) the PELOTON code, $f_e$ is the ionization fraction and $V$ is the discretized plasma volume the pellet is traveling through. $V$ is computed using tokamak specific data sets, and is considered static during the injection time period.

Electron and ion pressures evolve according to the following equations :
\begin{equation}
	\frac{d P_e}{d t} =  - Q_{{\cbl sink}} - Q_{rad} + n_e\nu_{ie}(T_i - T_e) + \frac{\gamma-1}{R^{2}}\eta(T_e) j^{2}
	\label{eq:Pe}
\end{equation} 

\begin{equation}
	\frac{d P_i}{d t} = - n_e\nu_{ie}(T_i - T_e)
	\label{eq:Pi}
\end{equation} 
where  $Q_{{\cbl sink}}$ {\cbl is the power density loss due to ionization, dissociation, and if plasmoid drift is accounted for, cloud heating by plasma electrons is included in this term}. $Q_{rad}$ is the radiation power density (for neon or mixed pellets). The last term in the right-hand side of equation (\ref{eq:Pe}) describes the Ohmic heating computed using the equilibrium current density profile, where the resistivity $\eta$ is computed using expressions from the NRL Plasma Formulary \cite{NRL}. The last term in the right-hand side of equation (\ref{eq:Pi}) describes the ion-electron collisional energy exchange and $\nu_{ie}$ is the electron–ion collision rate. 
We assume an ideal gas equation of state and find new background temperatures as
\begin{equation}
	T_e = \frac{P_e}{n_e k}, \quad T_i = \frac{P_i}{n_i k}
	\label{eq:Equation of State}
\end{equation} 

A mapping $\psi(R,Z)$ associates geometric (R,Z) coordinates in the poloidal plane with a poloidal flux value. Ablated material is deposited in discrete plasma volumes according to this mapping.

\section{Numerical implementation}

The system of equations (\ref{eq:mass}-\ref{eq:eos}) together with the pellet surface  ablation boundary conditions are implemented in PELOTON, a pellet / SPI ablation code
based on the Lagrangian particle (LP) method \cite{SamulyakWang2018} for hydrodynamic-type equations. The first version of PELOTON was introduced in \cite{SamulyakYuan2021} as a Lagrangian particle pellet code. PELOTON is applicable to pellets and SPI fragments ablated by hot plasma and runaway electrons \cite{YuanNaitlho2022}.  
The LP method represents fluid cells with Lagrangian particles and is suitable for the simulation of highly nonuniform / multiphase flows. The particle method is strictly mass-conservative and naturally adaptive to density changes, a critically important property for 3D simulations of the ablation of pellets and, especially, SPI fragments.
The LP code is optimized for massively parallel supercomputers \cite{YuanZepeda2023} by using p4est ("parallel forest of K-trees" ) \cite{BursteddeWilcox2011}, a parallel library that implements a dynamic management of a collection of adaptive K-trees on distributed memory supercomputers. 
PELOTON was validated using experimental data on the injection of hydrogen fueling pellets and the DIII-D experiment operating with small neon pellets \cite{HollmannNaitlho2022}. 

In present work, we implemented an improved non-uniform pellet cloud charging model of \cite{ParksLu2009} that is critical for rocket acceleration. Previous PELOTON simulations \cite{SamulyakYuan2021,HollmannNaitlho2022} assumed a constant electrostatic shielding potential of the ablation cloud. Adopting a non-uniform distribution of the shielding potential $\Phi$ in a transverse plane to the magnetic field leads to increased values of $n_{\rm eff}$ in the HFS compared to the LFS (see Figure \ref{fig1}) and the corresponding heat depositions and cloud pressures. The implementation involves solving equations (\ref{eq:phi} - \ref{eq:mu}) for every magnetic field line (i.e. a longitudinal integration line). The system of integration lines is constructed adaptively in such a way that their density is high in the vicinity of the pellet or SPI fragment and it decreases away from the pellet in transverse direction, reducing computational cost. The corresponding numerical tool, a parallel construction and refinement of an octree data structure containing all Lagrangian particles, is described in detail in \cite{SamulyakYuan2021}. We would like to note that the non-uniform electrostatic shielding model merely re-distributes the incoming hot electron flux between the HFS and the LFS, and leads to only negligibly small changes in pellet ablation rates compared to previous simulations with constant $\Phi$.

\section{Simulation results}

\subsection{Simulations at conditions of JET SPI and scaling law}

We performed a series of PELOTON simulations of the ablation of deuterium pellets and computed their rocket acceleration in a parameter range of plasma temperature and density and pellet radii pertinent to the deuterium SPI injection experiments in JET \cite{KongNardon2024}. As our purpose was to estimate a scaling law for the pressure difference $\Delta P$ on opposite sides of the pellet surface, responsible for the rocket acceleration, we ignored the influence of plasma gradients on $\Delta P$. Including such gradients would increase the parameter space dimension, requiring a significantly larger number of simulations for building a numerical database. We estimated the effect of plasma gradients in several stand-alone simulations, and concluded that in the SPI scenario, they affect $\Delta P$ during very short intervals of pellet fragment lifetimes. Due to the SPI injection direction (from the top, described in the following sections), plasma gradients typically change the rocket acceleration component in the propagation direction and not in the $\nabla R$ direction. Some estimates of plasma gradients for an ITER scenario are presented in Section \ref{sec:gradients}.

By performing a log-linear regression analysis of simulation data, we obtained the following scaling law for the rocket acceleration pressure difference $\Delta P$ in JET:  
\begin{equation}
	 \Delta P[bar] = 1.0  \left(\frac{T_e[keV]}2\right)^{1.26} \left(\frac{n_e[1/m^3]}{1e20}\right)^{0.4}.
	 \label{scaling_law}
\end{equation}
We used 20 simulation data points to determine this fit which covered the approximate parameter ranges $T_e \in $ [450 eV, 2000 eV], $n_e \in $ [6e19 $m^-3$, 3e20 $m^-3$], $r_{p}\in $ [0.2 mm, 1.8 mm], where $r_p$ is the pellet radius. {\cbl Exact simulation parameters and results are detailed in table \ref{table:JET_sim_data} within the appendix. } Comparison of this fit and data points computed by PELOTON gives the mean absolute relative error of approximately 13 percent. This error is mainly due to  numerical noise in computed rocket acceleration values because the mean signed relative error is only one percent. Therefore, we believe that (\ref{scaling_law}) accurately represents $\Delta P$ values for the given parameter range.
We observed that the rocket acceleration related pressure difference is practically independent on the pellet radius: the regression fit into numerical data yielded the following relation
\[
\Delta P \sim \left(\frac{r_p[mm]}2\right)^ {0.08} 
\]
which was ignored as it is within statistical error of numerical simulations. 
This observation was also confirmed in a separate study of SPI in ITER to be reported in a separate publication. This is likely caused by similarity of shapes and properties of the corresponding ablation clouds. We will study this phenomenon in more detail in our future work.
All simulations were performed using values of the magnetic field (3 T)  and the magnetic field radius of curvature (3 m) relevant to JET SPI experiments. Since the rocket acceleration strongly depends on the ablation cloud grad-$\mathbf{B}$ drift and the corresponding cloud asymmetry, the reported scaling law is not accurate for machines operating in a significantly different parameter space. The corresponding scaling law for ITER SPI fragments, injected from the outside along major radius, will be reported in a forthcoming paper.  

\subsection{Validation with deuterium SPI experiment in JET}

\label{sec: JET SPI Simulations}

We conducted deuterium SPI simulations in JET to demonstrate the impact of PELOTON calculated rocket acceleration on SPI trajectories. 
Using PELOTON's plasma cooling model, we dynamically ablate SPI fragments according to a scaling law consistent with PELOTON data, deposit ablated material on magnetic surfaces, compute the corresponding change of the background plasma density and temperature, apply the pressure difference scaling law (\ref{scaling_law}) together with equation (\ref{eq:acceleration}) to dynamically compute rocket acceleration of SPI fragments and evolve their trajectories accordingly. Some aspects of this process are described in more detail below, followed by discussion of results.

First, we form initial spatial and velocity distributions of SPI fragments being injected.  We constrain large fragments to speeds near the mean value following \cite{GebhartBaylor2021} where they find that only about one percent of mass is found at the front and back of the SPI plume. We inject large fragments near the right side of the cone following images in \cite{GebhartBaylor2020} indicating that large fragments tend to be least deflected by the bend in the shatter tube. We develop a method of assigning speed and injection angle to fragments based on radius to achieve a plume with such characteristics.

\begin{figure}[h!]
\centering
\subfigure[Fragment speeds when entering plasma]{\includegraphics[width=.75\linewidth]{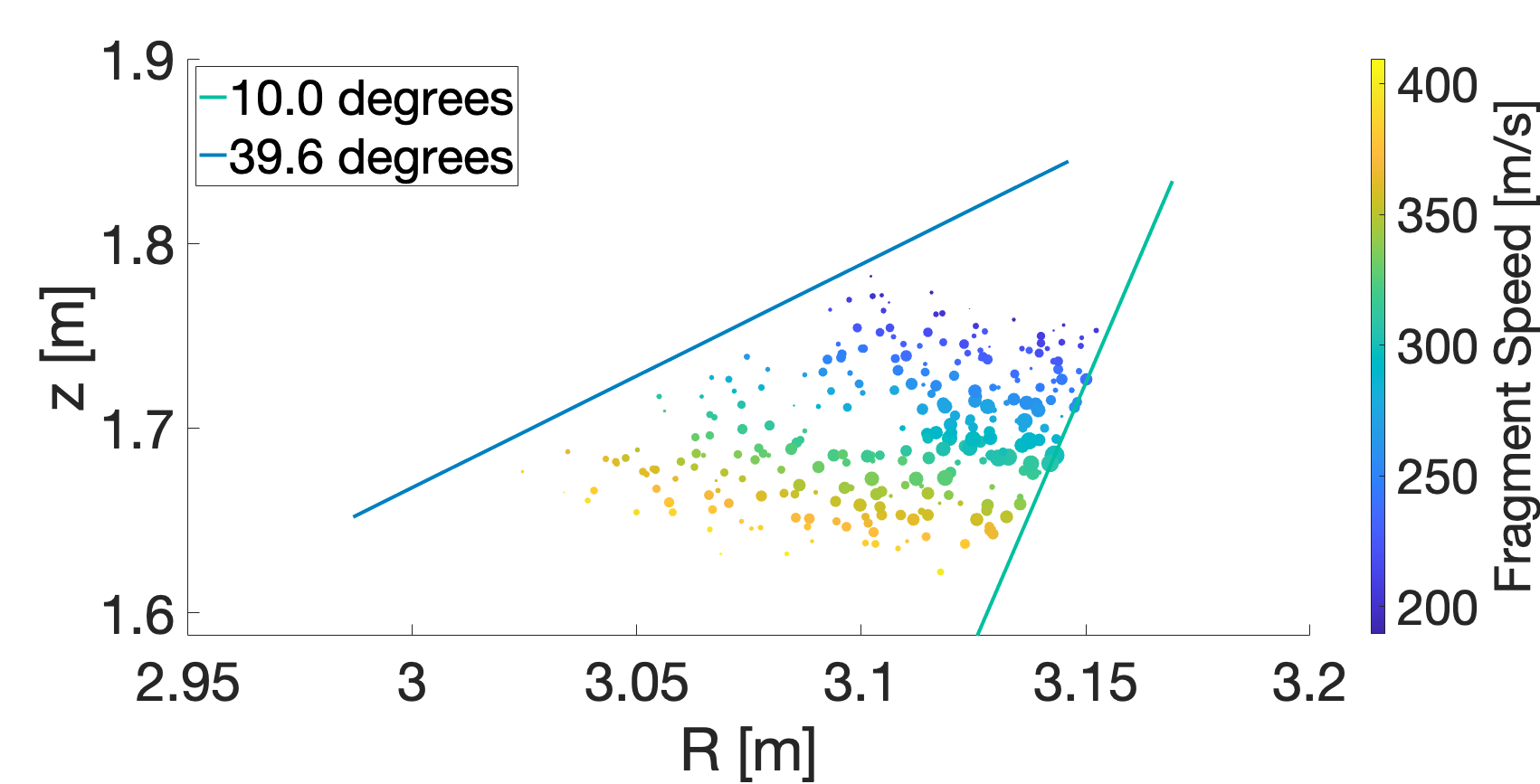}}
\subfigure[Fragment radii when entering plasma]
    {\includegraphics[width=.75\linewidth]{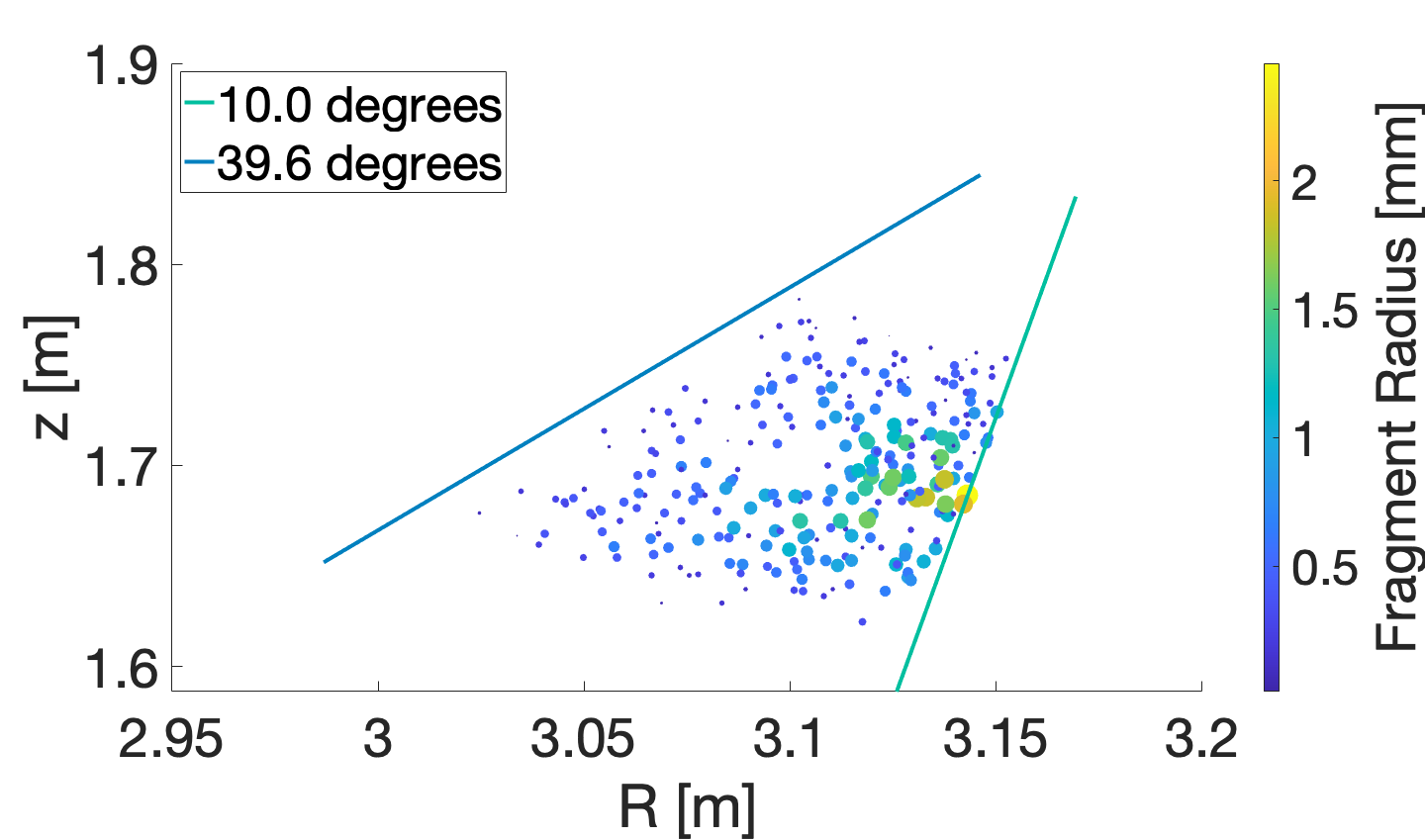}}
    \caption{SPI plume when fragments enter the plasma. Color mappings provide the speed (a) and radius (b) of each fragment. Additionally, fragment size corresponds to marker size.  The plume boundaries are marked by solid lines with angles measured clockwise from the downward vertical direction centered at the injection location in the poloidal plane.}
	\label{spi_characterization_plume}
\end{figure}

We use the same fragment size distribution as in \cite{KongNardon2024}, which is based on the fragmentation model presented in \cite{GebhartBaylor2020}. SPI speeds are determined in a similar manner to \cite{KongNardon2024}, from an interval with mean 300 m/s and width 40 percent of the mean, but we include bias related to the size of fragments. Fragment speeds are uniformly sampled from the interval [$180 + \Delta v$, $420 - \Delta v$], where $\Delta v = r*(0.4*300)/(r_{max})$, r is the fragment radius and $r_{max} = 2.05$ mm. 
Following images in \cite{KongNardon2024}, we confine all fragments to the interval of angles [$10.0^\circ$, $39.6^\circ$] as measured clockwise from the downward vertical direction centered at the injection location in the poloidal plane. The angle of injection for a fragment with radius r is uniformly sampled from the interval [$10.0^\circ$, $39.6^\circ - \Delta \theta$], where $\Delta \theta = r*(29.6^\circ)/(r_{max})$. Large fragments with radius near $r_{max} $ have an injection angle near the largest possible angle, the right-hand side of the injection plume in figure \ref{spi_trajectories}(a), the smallest angles can only be assigned to the smallest fragments. The resulting distributions of size, speed, and angle are characterized by Figure \ref{spi_characterization_plume}.


During each fragment ablation, the ablated plasmoids expand longitudinally along magnetic field lines, filling the flux volumes, and experience drift in the $\nabla R$ direction. Due to plasmoid drift, the location of effective material deposition influencing the ambient plasma states is shifted with respect to the plasmoid location. This significantly changes plasma profiles and the resulting ablation dynamics of fragments. {\cbl We adopt a `teleportation' model similar to \cite{KongNardon2024}, whereby the effects of plasmoid drift are approximated by depositing material locally around a location shifted in the direction of the plasmoid drift from the ablation location. We adopt this model as an approximation and do not explicitly track the material deposition during the plasmoid drift process.} 
A fixed 40 cm shift was used in \cite{KongNardon2024} to match experimental measurements. The actual plasmoid drift, however, depends on the states of plasmoids and the ambient plasma \cite{VallhagenPusztai2022,VallhagenAntonsson2025}. In this work, we adopt a variable plasmoid shift ad hoc model approximating \cite{VallhagenAntonsson2025}. Assuming  that the plasmoid drift is generally proportional to fragment's ablation rate, we apply $R_{shift} = \alpha*G$, where G is the ablation rate and $\alpha$ is an adjustable parameter. We use a value of $\alpha = 20 \frac{m*s}{kg}$ as it results in a reasonable agreement in the overall penetration depth of SPI fragments with experiments, {\cbl as fragment trajectories reach comparable depth as the curve indicative of the cloud edge in terms of the flux coordinate }. In addition, the energy deposition accounts for the energy gained by ablated material due to dissociation, ionization, and heating while drifting through different parts of the plasma. {\cbl We adopt the approximation that all such energy is provided to the ablation cloud by the region local to the ablating fragment and do not explicitly track energy gain during the drift process}. This energy (100 eV per ablated atom) is deducted from the location of ablating fragment and is added at the deposition location. 

\begin{figure}[h!]
	\centering
	\subfigure[]{\includegraphics[width=.49\linewidth]{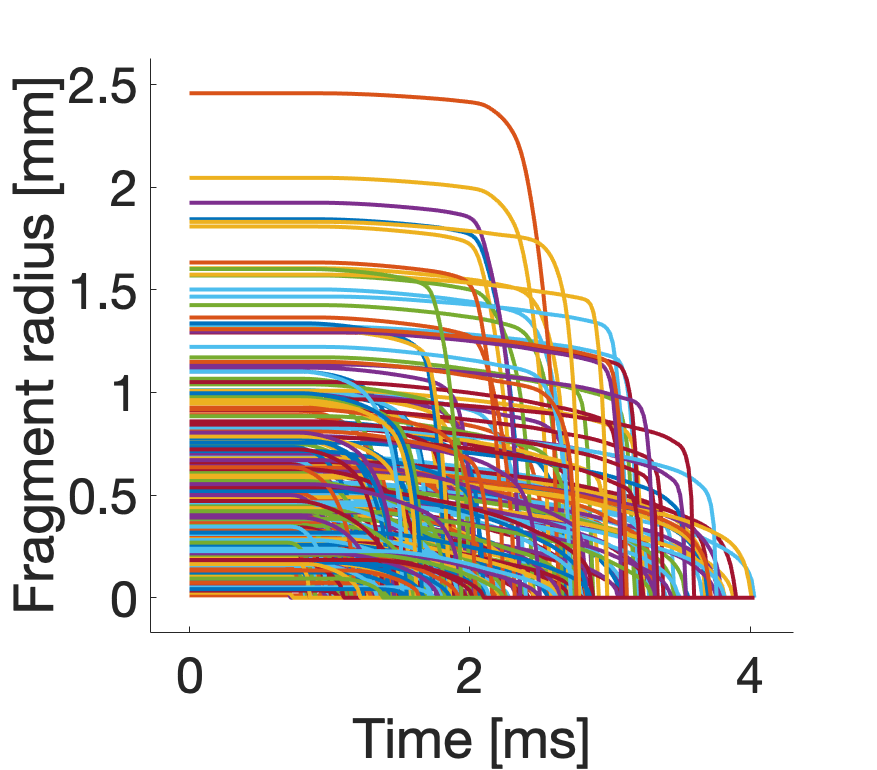}}
	\subfigure[]{\includegraphics[width=.49\linewidth]{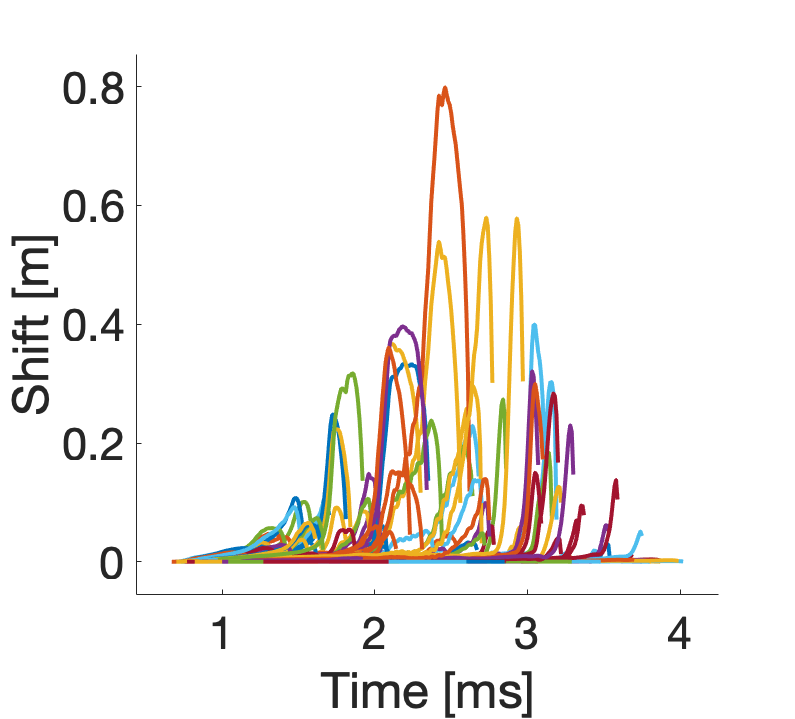}}
	\subfigure[]{\includegraphics[width=.49\linewidth]{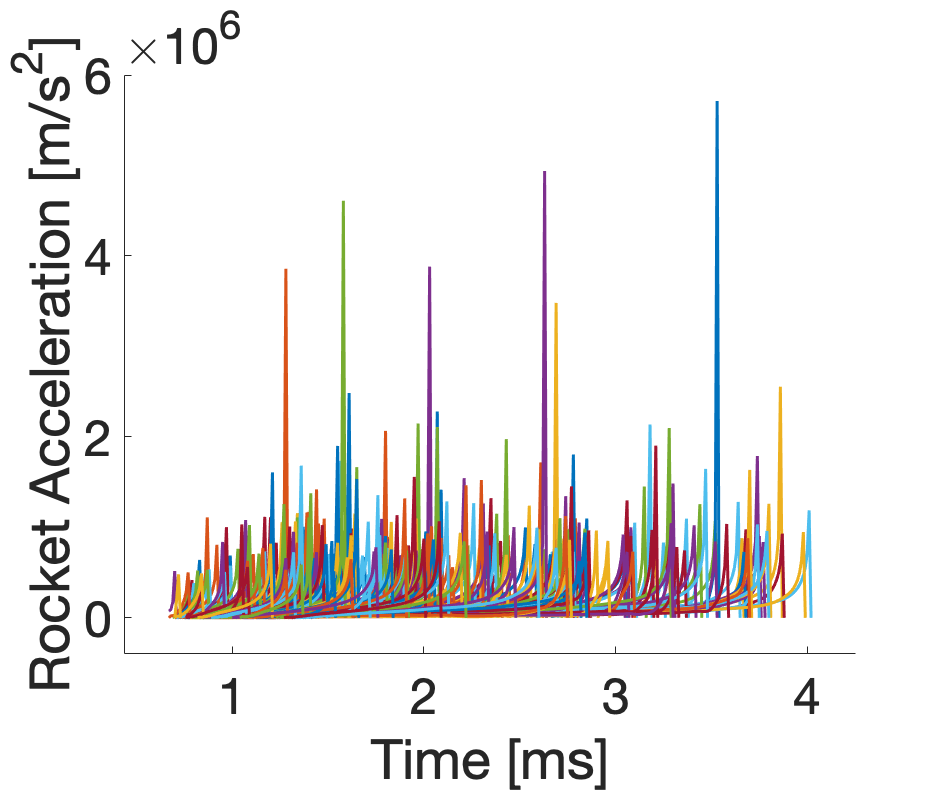}}
	\caption{(a) Evolution of fragment radii over the simulation time. (b) Evolution of the $R_{shift}$ used in ad hoc model for deposition of ablated material. (c) Evolution of rocket acceleration experienced by each fragment as a function of time.  Colors are consistent among three plots.}
	\label{spi_ablation}
\end{figure}

\begin{figure}[h!]
	\centering
	\subfigure[]{\includegraphics[width=.50\linewidth]{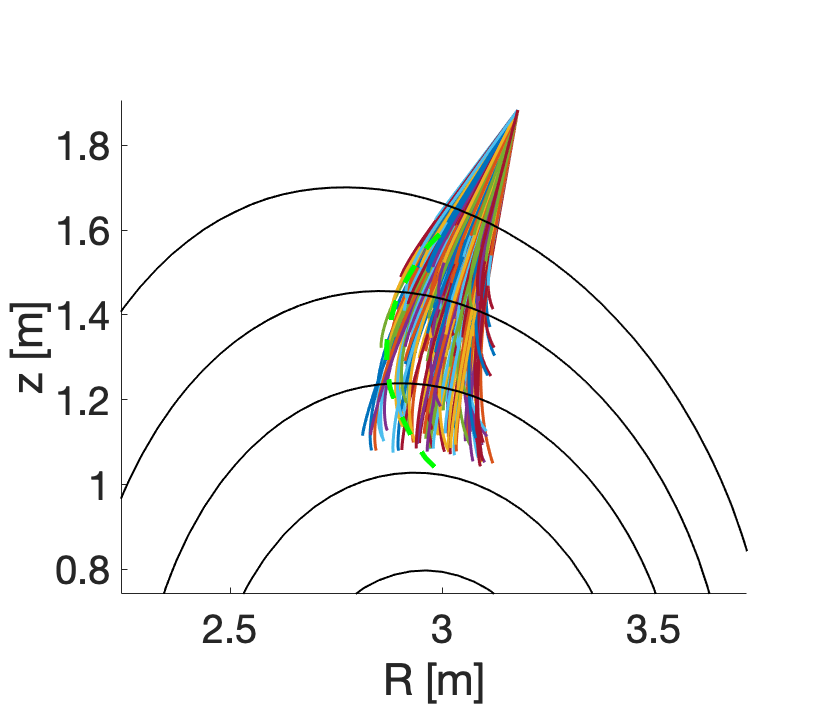}}
	\subfigure[]{\includegraphics[width=.48\linewidth]{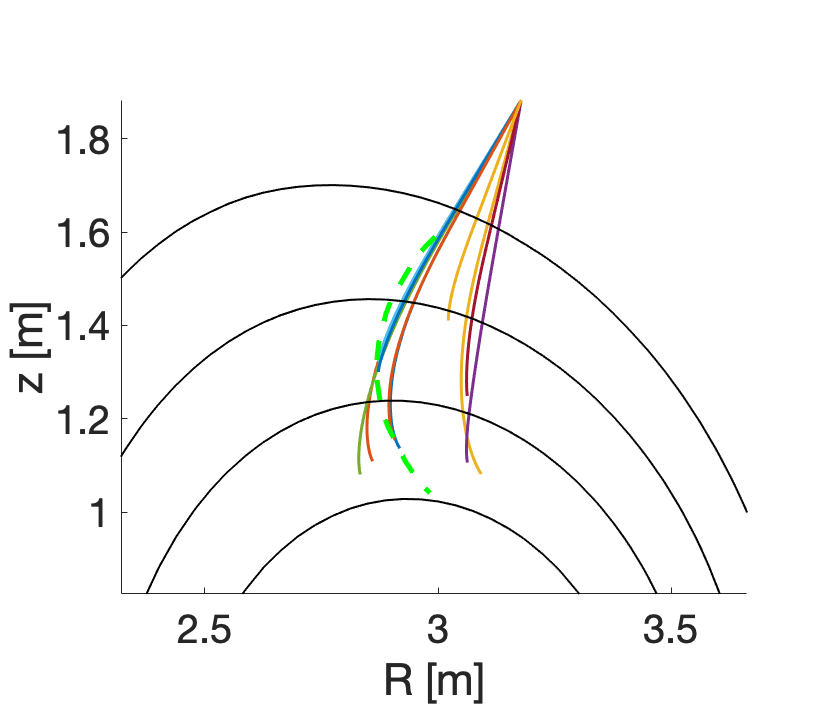}}
	\caption{(a) Trajectories of all SPI fragments during the simulation. (b) Trajectories of selected fragments. In both cases, the green dashed line represents the tracked inboard edge of the SPI plume based on the maximum gradient of the $D\alpha$ emission intensity detected by the KL8 camera in JET \#96874 \cite{KongNardon2024}.}
	\label{spi_trajectories}
\end{figure}

Figure \ref{spi_ablation}(a) shows the evolution of fragment radii. The overall ablation process lasts 4 ms.  Figure \ref{spi_ablation}(b) shows drift induced shifts experiencing by the corresponding plasmoids. Colors in Figure \ref{spi_ablation} are consistent among three plots showing, for example, that the largest fragment experienced the shift  of 0.8 m at the peak of its ablation rate. {\cbl In such cases, the plasmoid drifts into the SOL. From the perspective of the material deposition in the plasma, such material and the energy gained during drift is not added to the plasma and is effectively discarded from the simulation.}  The evolution of rocket acceleration values experienced by all fragments are depicted in  \ref{spi_ablation}(c). As expected, the rocket acceleration rapidly increased during short intervals of time at the end of each fragment lifetime due to decreasing mass, and pressure difference independent of fragment radius. Rocket acceleration in this simulation reached  maximum values greater than $5\times 10^6\, m/s^2$. Typical rocket acceleration values remained below  $10^6\, m/s^2$ for most of fragment lifetimes.

Figure \ref{spi_trajectories}(a) shows computed trajectories of all SPI fragments in the considered JET case. The green dashed line displays the experimentally measured edge of the SPI plume based on the maximum gradient of the $D\alpha$ emission intensity detected by the KL8 camera in JET \#96874 \cite{KongNardon2024}. The shape of the region occupied by computed SPI trajectories and their deviation to the $\nabla R$ direction are consistent with experimental observations reported in   \cite{KongNardon2024}.

\begin{figure}[h!]
	\centering
	\includegraphics[width=.55\linewidth]{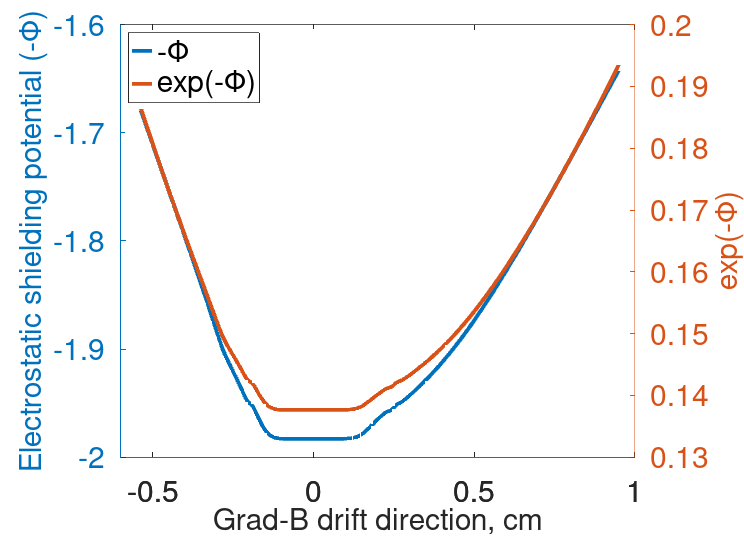}
	\caption{\cbl Electrostatic shielding potential across the ablation cloud in the direction of grad-B drift (blue line) and reduction of the incoming hot electron density by electrostatic shielding (red line)}	
	\label{phi_shielding}
\end{figure}

{\cbl To clarify the influence of various factors on the rocket acceleration, we plotted the electrostatic shielding potential across the ablated cloud and the corresponding reduction of the  incoming hot electron density $\exp(-\Phi)$ (blue and red lines in Figure \ref{phi_shielding}, correspondingly). As expected, the potential is steeper on the HFS, providing less shielding at the same distance from the pellet compared to the LFS. However, the electrostatic shielding potential effect on the rocket acceleration is much smaller compared to the difference of decay of hot electrons on both sides of the pellet due to the ablation cloud asymmetry. We estimated that the energy deposition per unit mass into ablated plasma at 1 mm off the pellet surface was about 35\% higher on the HFS  compared to the LFS. The corresponding difference in the incoming hot electron reduction $\exp(-\Phi)$  was only 1.4\%.  We also verified that the low magnetic Reynolds approximation is accurate for the denser part of the plasmoid defining the ablation rate and rocket acceleration. 
}

\begin{figure}[h!]
	\centering
	\subfigure[]{\includegraphics[width=.49\linewidth]{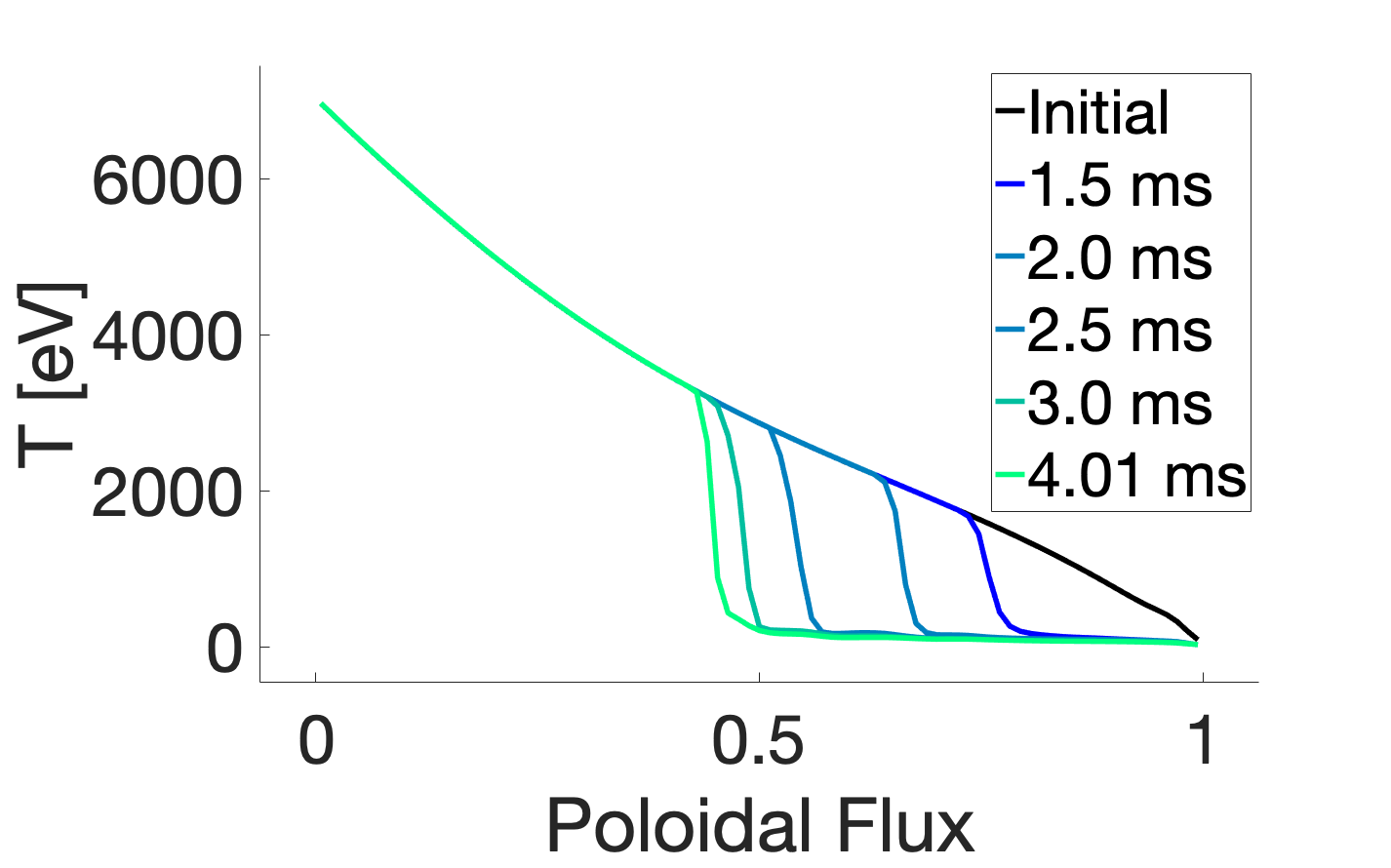}}
	\subfigure[]{\includegraphics[width=.49\linewidth]{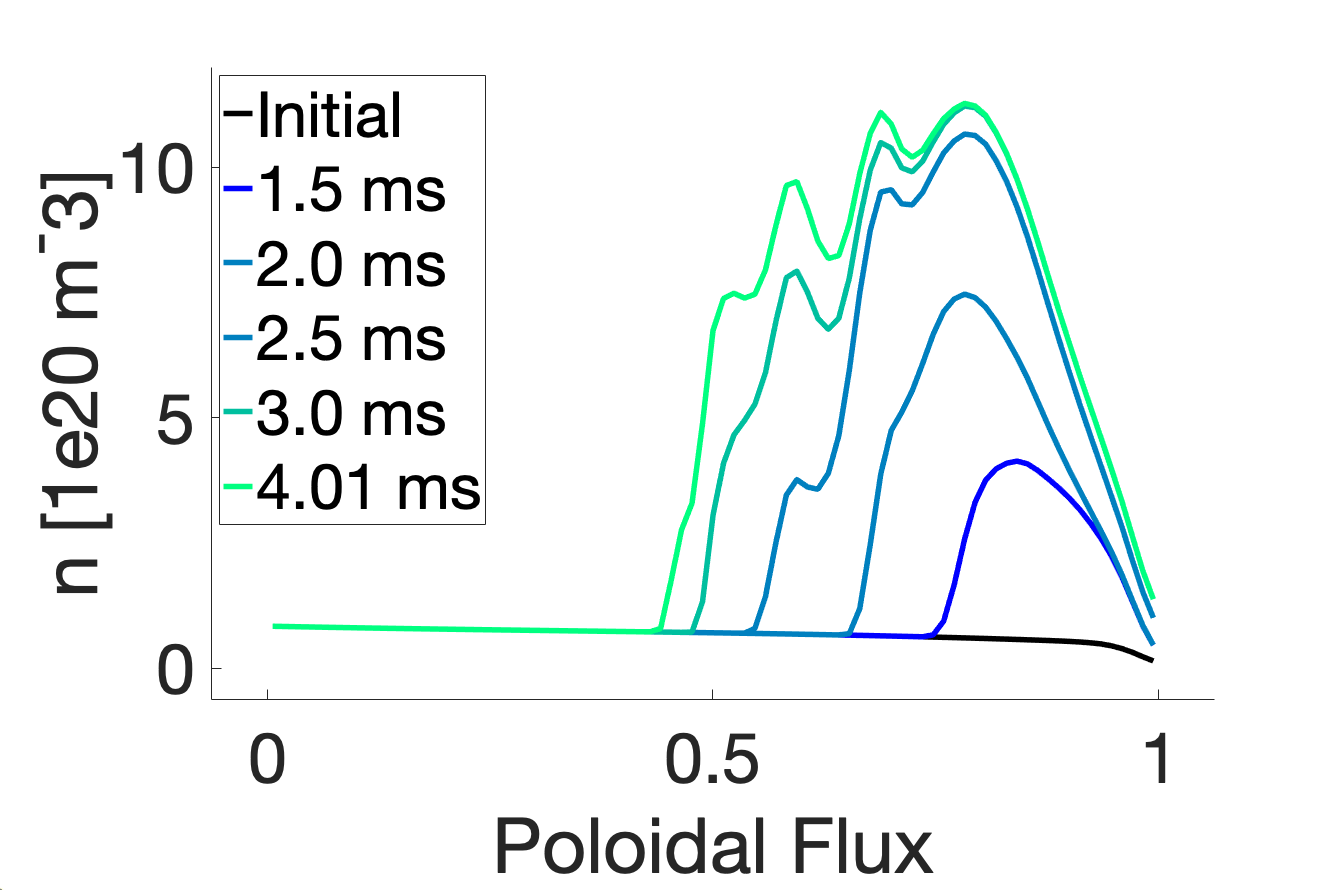}}
	\caption{(a) Electron temperature and (b) electron density predicted by PELOTON's cooling module at selected times. }
	\label{Plasma_profiles}
\end{figure}

\begin{figure}[h!]
	\centering
	\subfigure[]{\includegraphics[width=.49\linewidth]{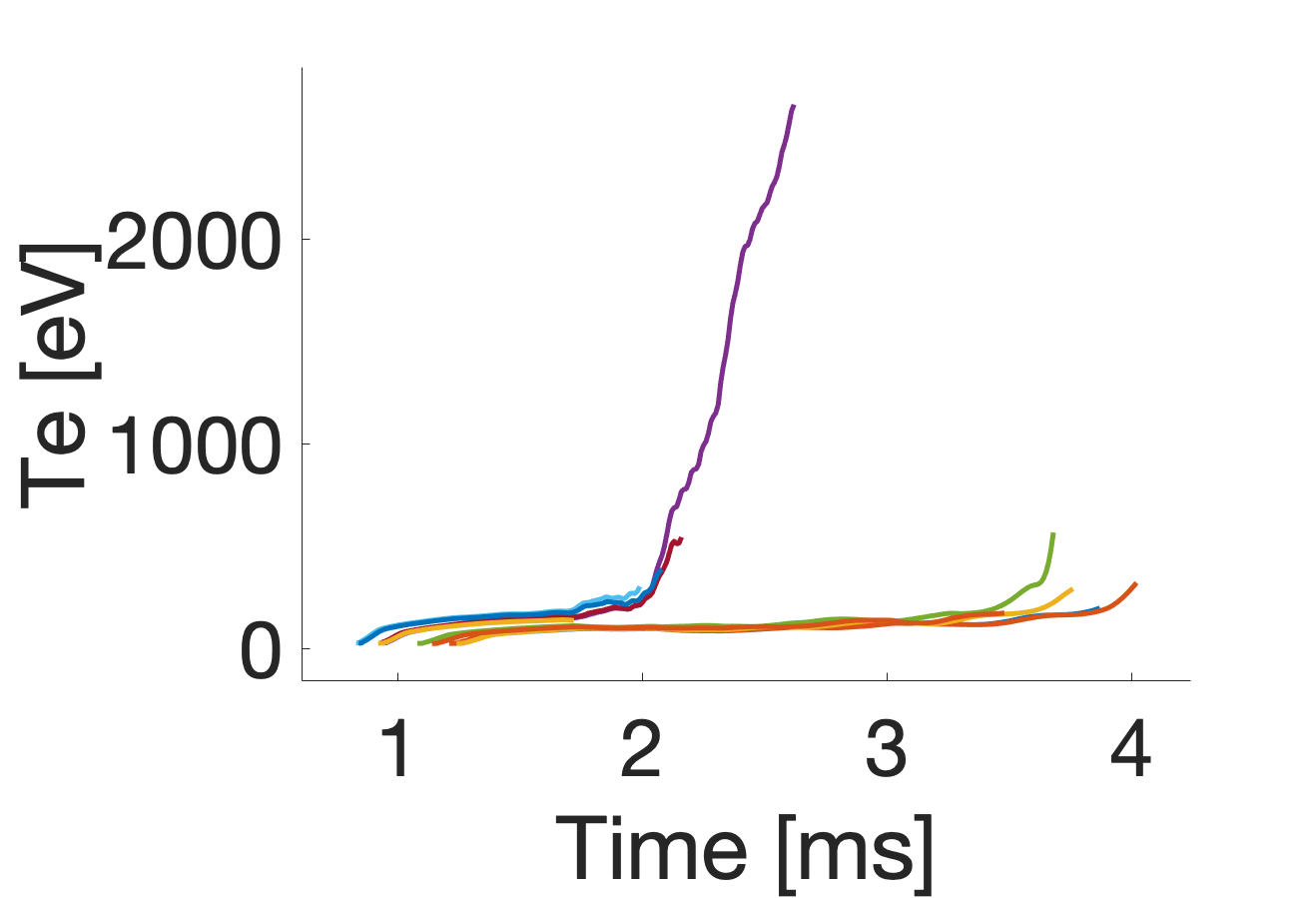}}
	\subfigure[]{\includegraphics[width=.49\linewidth]{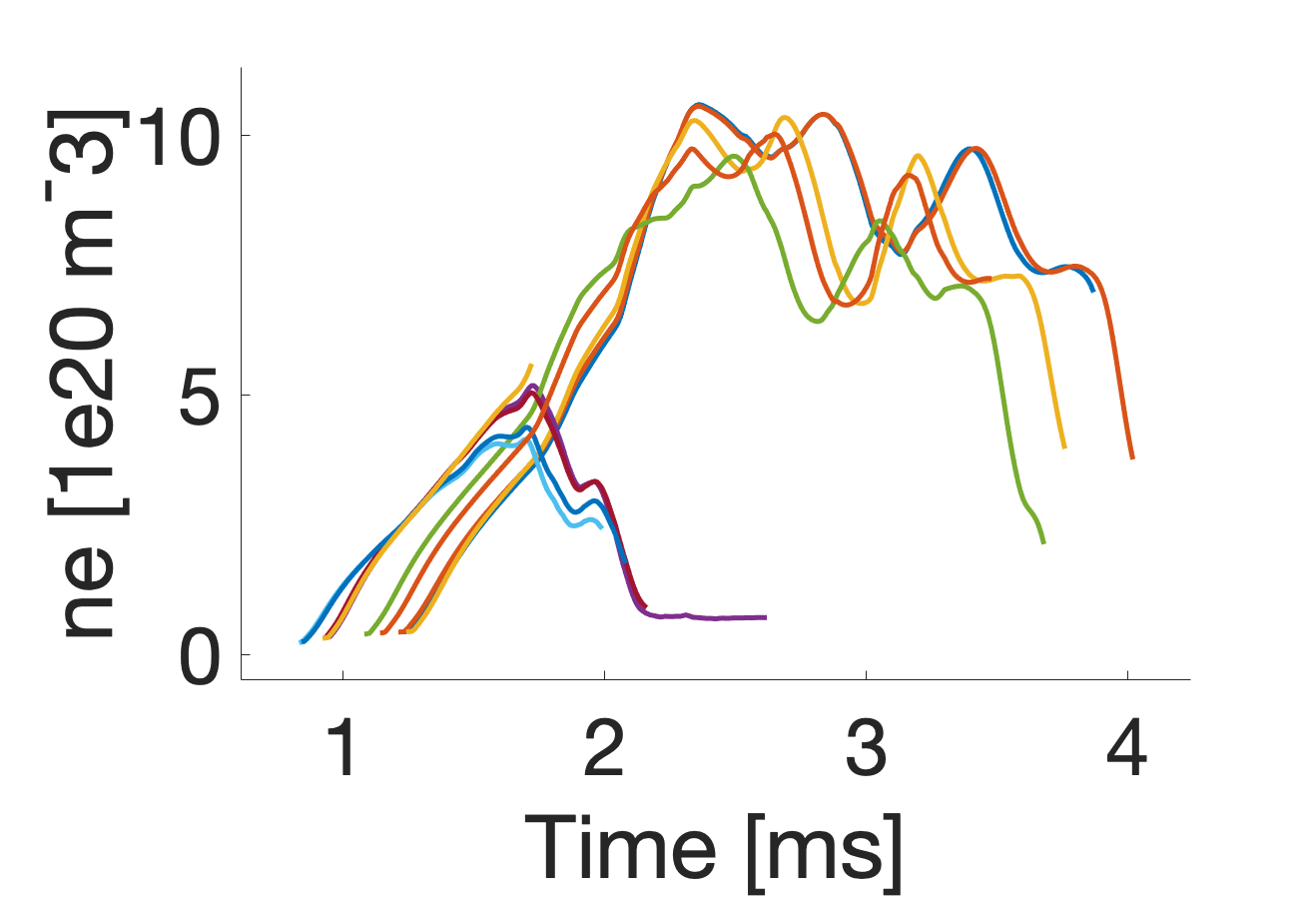}}
	\caption{(a) Temperature experienced by selected fragments depicted in Figure \ref{spi_trajectories} (b) as a function of time. b) Density experienced by selected fragments depicted in Figure \ref{spi_trajectories} (b) as a function of time. }
	\label{Trajectory_states}
\end{figure}

Finally, we discuss the plasma profiles obtained by PELOTON's dilution cooling code. Plasma profiles in JET at selected times are displayed in Figure \ref{Plasma_profiles}. {\cbl We observe the increase of plasma density and reduction of temperature with the deposition of new ablated material.} 
For selected SPI fragment trajectories shown in Figure \ref{spi_trajectories}(b), electron temperature and density experienced by each fragment as a function of time is shown in figure \ref{Trajectory_states}. {\cbl Most of selected fragments ablated at low plasma $T_e$ and high $n_e$. Despite reduced values of their rocket acceleration compared to the fragment that experienced plasma $T_e$ above 2 keV (purple line), trajectories of these fragments deviated more from the straight line because of their long lifetime / duration of the rocket acceleration. The exception was the fragment with shortest penetration depth due to small initial size, plotted with a yellow line (see Figure \ref{spi_trajectories}(b)). }

PELOTON's cooling code over-predicts density at the plasma edge when the shift model for the ablated material deposition of \cite{KongNardon2024} is applied to model the effect of the plasmoid drift. PELOTON's cooling module does not include diffusion across flux surfaces and MHD effects. We observe density gradients near the plasma edge caused by shifted deposition of material and we attribute this to the missing physics effects that would diffuse plasma particles across the scrape-off-layer (SOL). We performed tests designed to explore the impact of lower densities at the plasma edge and corresponding higher temperatures on SPI fragment trajectories. We artificially reduced densities when specified cut-off levels are reached and correspondingly increased temperatures. We did not display results for conciseness, but observed shorter fragment lifetimes, particularly among smaller fragments. The overall qualitative behavior of the plume remained very similar. {\cbl The overall behavior is similar because of the shifted deposition of ablated material. The over-predicted density buildup at the plasma edge mostly impacts the relatively small fragments in the trailing end of the SPI plume.}

\subsection{Influence of neon component}

In this section, we examine the influence of 0.5\% neon addition (in terms of atomic ratio) to the pellet material on the rocket acceleration. Figure \ref{fig:d2_neon_traject} compares trajectories of a typical deuterium fragment with the initial radius of 1.9 mm, injected close to the middle of the SPI cone, with an identical deuterium-neon fragment in terms of the initial radius and injection velocity. Both fragments use exactly the same plasma states predicted by the PELOTON cooling module. By ignoring the effect of radiation cooling of  the ambient plasma in the presence of neon SPI, we aim to observe the change of rocket acceleration due to different properties of the corresponding ablation clouds. 

\begin{figure}[h!]
	\centering
	\includegraphics[width=.5\linewidth]{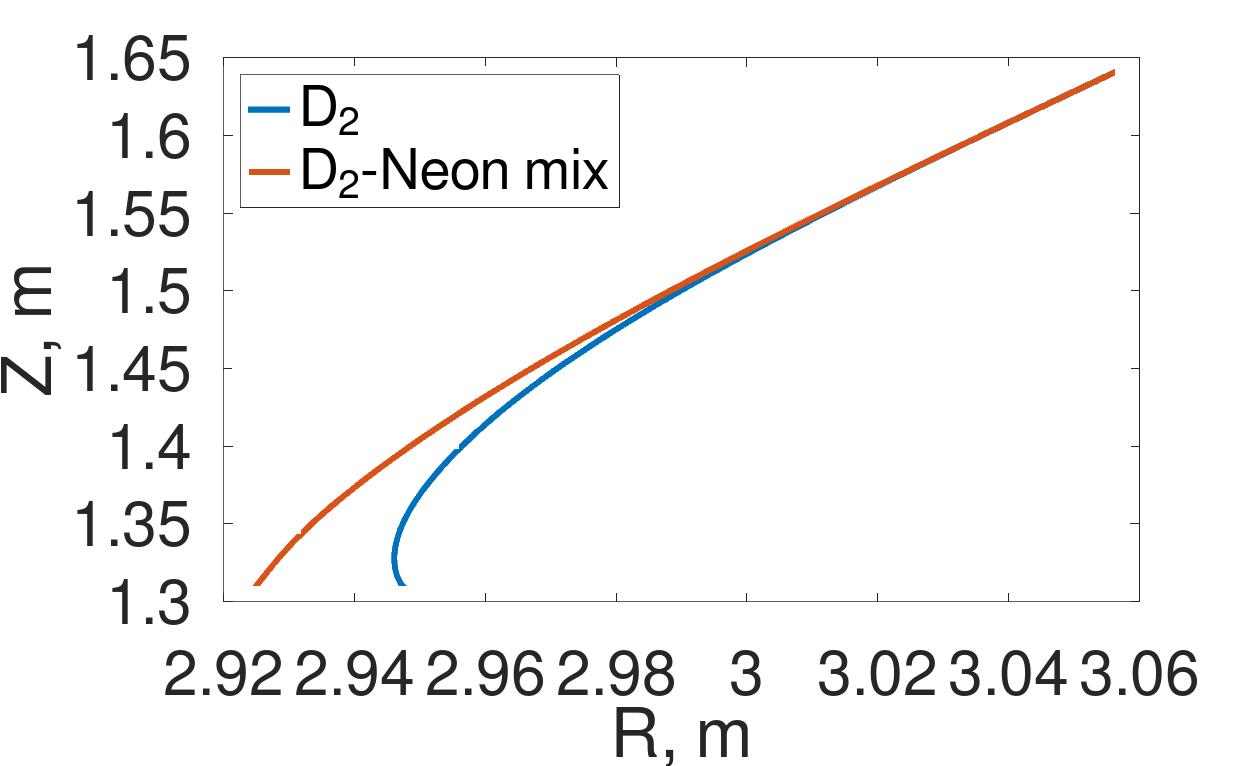}\label{fig:d2_neon_traject}
	\caption{Comparison of trajectories for 1.9 mm radius fragments composed of pure deuterium and of deuterium-neon mixture. The blue line shows the trajectory for the pure deuterium fragment and the orange line depicts the trajectory for the fragment containing deuterium-neon mixture with 0.5\% atomic ratio of neon. Both trajectories are shown from the plasma edge.}
	\label{fig:d2_neon_traject}
\end{figure}

\begin{figure}[h!]
	\centering
	\subfigure[pure deuterium]{\includegraphics[width=.9\linewidth]{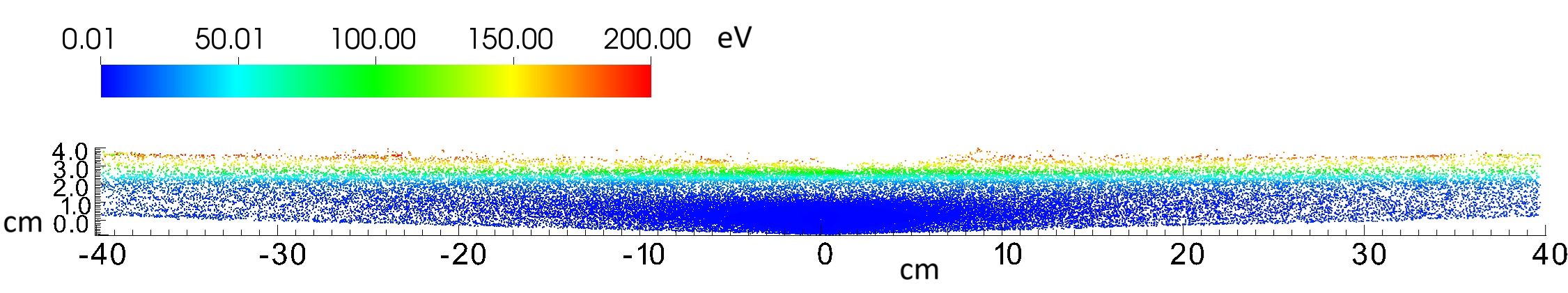}}
	\subfigure[deuterium-neon]{\includegraphics[width=.9\linewidth]{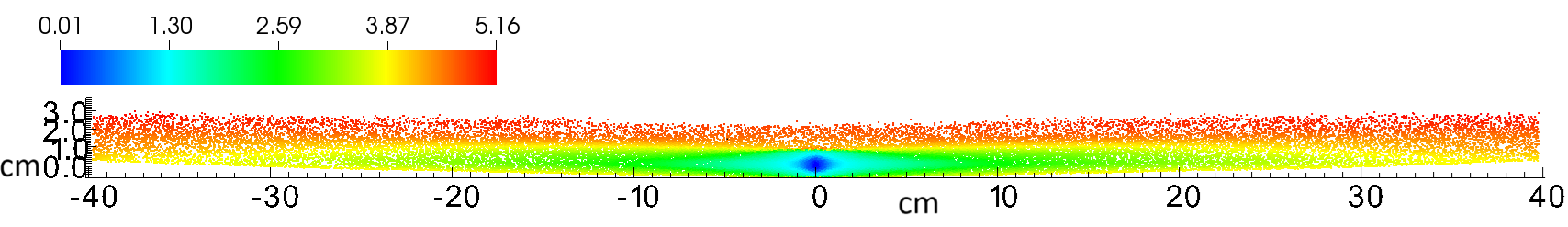}}
	\caption{ Distributions of temperature (eV) on thin slices of ablation clouds oriented in the directions of the magnetic field and the grad-$\mathbf{B}$ drift are shown for the pure deuterium fragment and the fragment of deuterium-neon mixture containing 0.5\% of neon.
    The clouds represent simulations of the ablation cloud at a selected instant along the fragment's trajectory. At the selected point, the fragment radius is 0.8 mm and the background plasma state is $T_e = 1400$ eV and $n_e =  8.5e19 \, m^{-3}$.}
	\label{T_d2_neon}
\end{figure}

Figure \ref{T_d2_neon} depicts temperature distributions on cross-sections of ablations clouds created by pure deuterium and deuterium-neon fragments at some mid-point of their trajectories.  
 While the presence of neon significantly reduces temperature values in the ablation cloud compared to pure deuterium, it also reduces both the longitudinal expansion velocity and the grad-$\mathbf{B}$ drift velocity. As a result, the deuterium-neon mixture clouds preserve asymmetric shape, leading to slightly smaller shielding length / larger electron heating rate of ablated plasma on the HFS compared to the LFS, and the rocket acceleration remains noticeable. While it is significantly lower compared to pure deuterium (partially due to the increased fragment mass), it can not be neglected. 
  We attribute the absence of rocket acceleration effect on trajectories of deuterium-neon fragments in the corresponding SPI experiments in JET to significantly changed background plasma states due to the prompt triggering of a global MHD event when neon is present in the pellet \cite{JachmichLehnen2025} \cite{KongJachmich2025}.

{\cbl
\subsection{Influence of Magnetic Field Strength and Radius of Curvature}
	In this section we report results from a parameter scan about one simulation point at JET conditions. In particular, we select point 20 from table \ref{table:JET_sim_data} in the appendix and perform four simulations, each of which varies one parameter at a time. We vary the magnetic field strength B and the pellet's radial coordinate R. Simulation point 20 used B = 2.8 T and R = 3 m for a 1 mm pellet with background plasma with electron temperature and density at approximately 1220 eV and  1.25e20  $m^{-3}$, respectively. As such, varying one parameter at a time gives simulations at B = 3.8 T, B = 1.8 T, R = 4.0 m, and R = 2.0 m, where we have selected to vary B by 1 T and R by 1 m. The results are presented in tables \ref{table: R scan} and \ref{table: B scan}. Varying the R coordinate did not have a large impact on the ablation rate or the pressure asymmetry leading to the rocket acceleration. We observed that the ablation rate slightly decreased with increases in the magnetic field, and that the pressure asymmetry increased with increasing magnetic field strength.}

\begin{table}[H]
	
	\centering
	\begin{tabular}{|c|c|c|} 
		\hline
		Radial Coordinate (m) &G (g/s) & $\Delta P$ (bar) \\
		\hline
		4.0    &10.2  &0.70  \\
		3.0  &10.3  &0.68  \\
		2.0    &10.4  &0.6  \\
		\hline
	\end{tabular}
	\caption{{\cbl We perform simulations in which we vary the radial coordinate by 1 m, leaving all other simulation parameters the same as in point 20 in table \ref{table:JET_sim_data}. Point 20 is represented by the middle row. In particular, these simulations use a 1 mm pellet with background plasma at approximately electron temperature 1220 eV and electron density 1.25e20  $m^{-3}$}}
	\label{table: R scan}
\end{table}

\begin{table}[H]
	\centering
	\begin{tabular}{|c|c|c|}
		
		\hline
		
		B (T) &G (g/s) & $\Delta P$ (bar) \\
		\hline
		3.8  &9.9  &0.85  \\
		2.8  &10.3  &0.68  \\
		1.8  &10.95  &0.4  \\

		\hline
	\end{tabular}
	\caption{{\cbl We perform simulations in which we vary the magnetic field strength by 1 T, leaving all other simulation parameters the same as in point 20 in table \ref{table:JET_sim_data}. Point 20 is represented by the middle row.  In particular, these simulations use a 1 mm pellet with background plasma at approximately electron temperature 1220 eV and electron density 1.25e20  $m^{-3}$}}
	\label{table: B scan}
\end{table}

\subsection{Influence of plasma gradients and partial screening}

\subsubsection{Plasma gradients}
\label{sec:gradients}

In this section, we study the influence of plasma gradients on the rocket acceleration of SPI fragments. We demonstrate this effect by using pure hydrogen SPI injection into ITER H mode. SPI fragments were injected from the initial position R = 8.568 m, Z = 0.685 m, with velocities in the range from  300 m/s to  700 m/s, within a 20 degree cone along the major radius. Data sets from 15 MA DT H-mode scenario (DThmode24) from CORSICA scenarios \cite{KimPoli2017} at 400s were used.

\begin{figure}[h!]
	\centering
	\subfigure[Density]{\includegraphics[width=.49\linewidth]{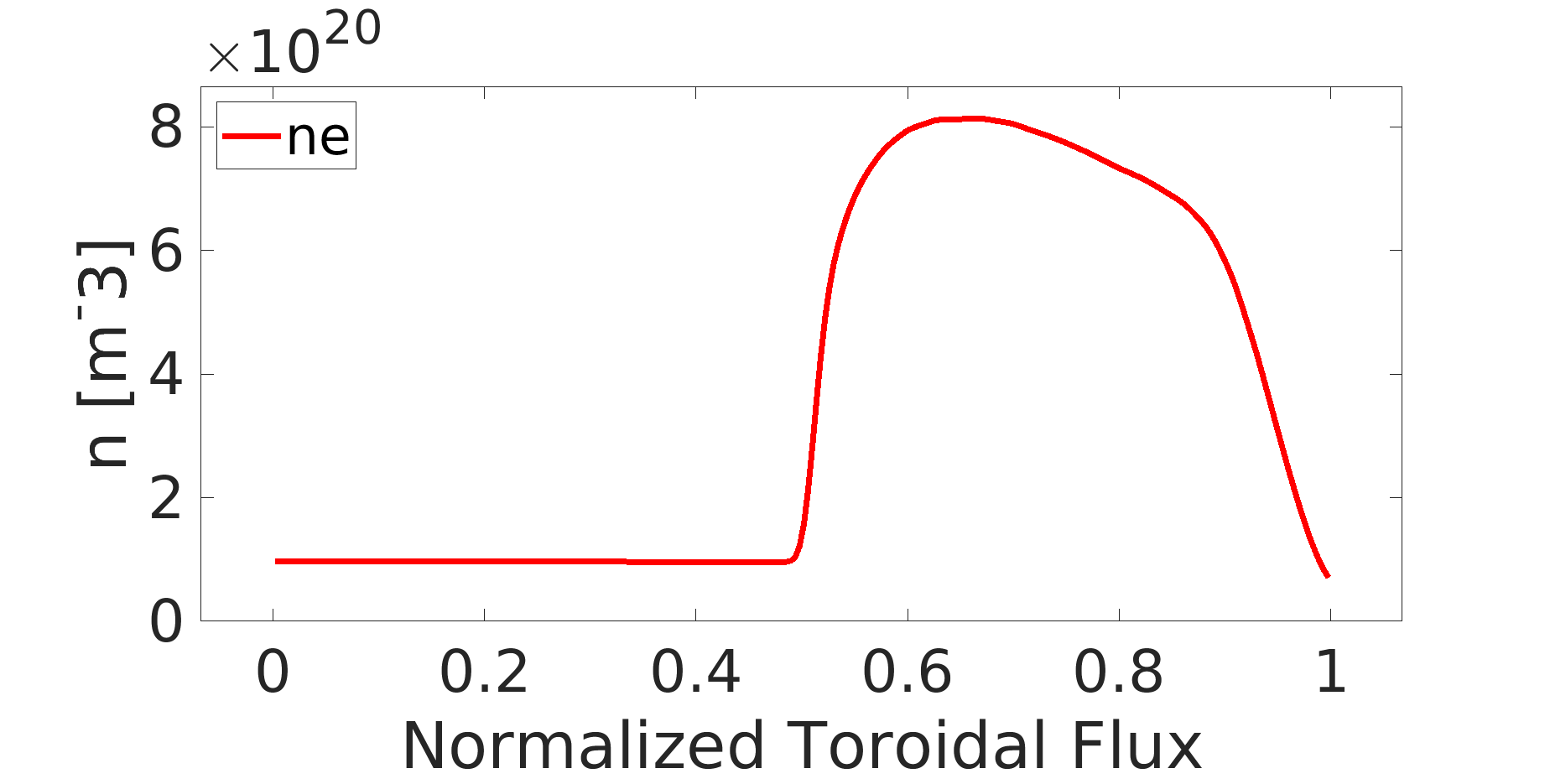}}
	\subfigure[Temperature]{\includegraphics[width=.49\linewidth]{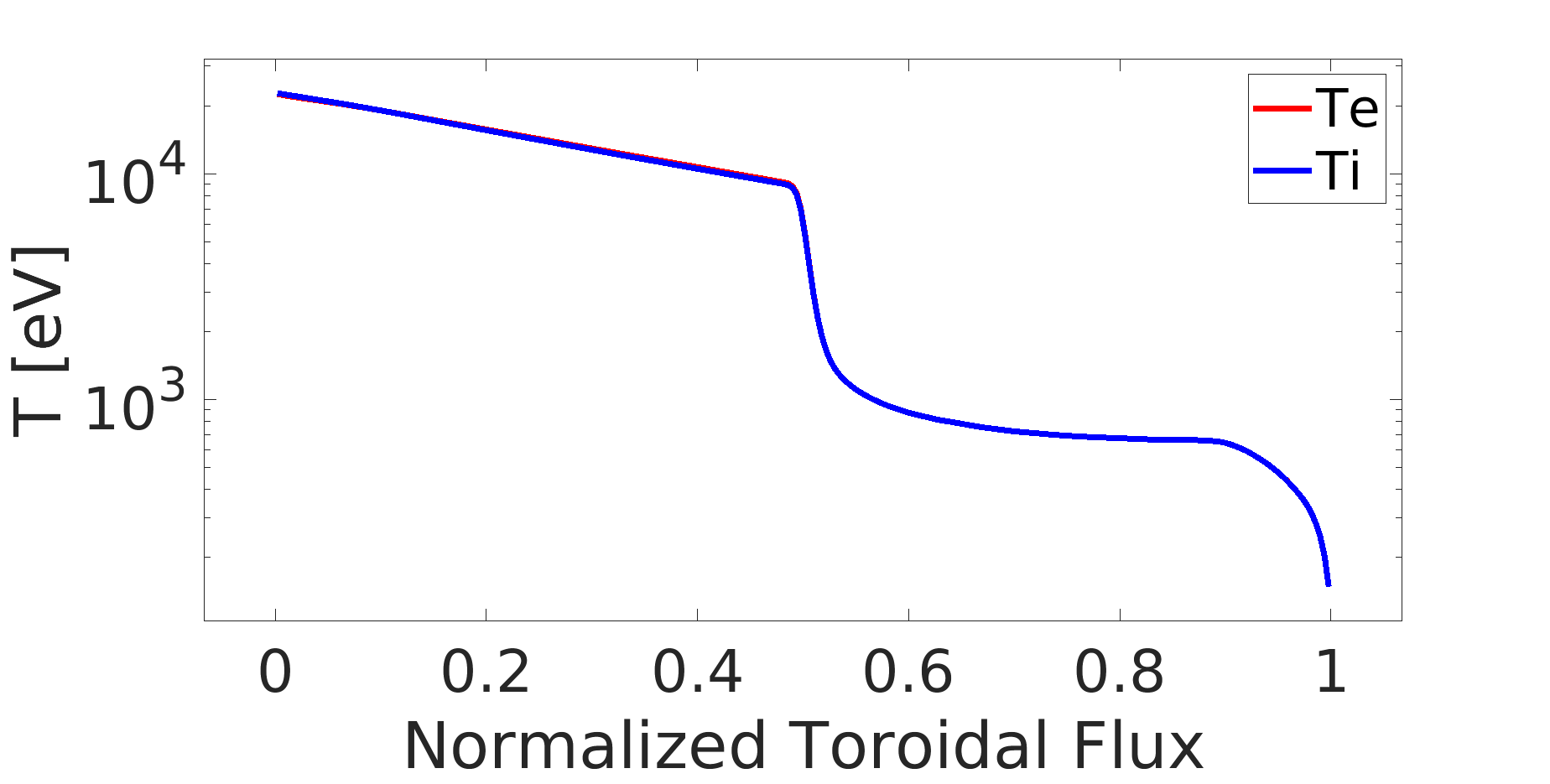}}
	\caption{(a) Plasma electron density, $m^{-3}$ and (b) temperature (eV) after the injection of 30 SPI fragments.}
	\label{fig:ITER_plasma_30SPI}
\end{figure}

Using PELOTON's plasma cooling module, we computed plasma states after the injection of first 30 fragments (see Figure \ref{fig:ITER_plasma_30SPI}), about 10 percent of a typical ITER injection, and used these states to estimate how plasma gradients affect the rocket acceleration. As we observe, after ablated material is deposited the plasma electron temperature remains practically constant, approximately 600 eV, within normalized flux coordinates from 0.9 to 0.5; plasma density also does not experience large gradients. As a result, plasma gradients changes the rocket acceleration by  negligibly small amounts in this domain. Therefore, we estimate the upper bound of the plasma gradient effect by considering largest plasma gradients at the edge of the ablated material deposition with the flux coordinate of about 0.5.

For the demonstration, we selected a 1.3 mm fragment. The plasma states and computed rocket acceleration values are shown in Table \ref{table1}. We observe that indeed, plasma gradients significantly change the rocket acceleration, increasing it by a factor of 3. Nevertheless, such a large acceleration is acting during a very short time interval as the fragment crosses the ablated material deposition boundary.  In addition, as we mentioned before, the plasma gradients are slightly overestimated in our cooling module due to absence of proper diffusion physics.

\begin{table}[h!]
	\centering
	\begin{tabular}{|c|c|c|c|c|c|c|}
		\hline
		$T_e$, eV & $n_e, \, m^{-3}$ & $B$, T & $\partial T/\partial R$ & $\partial n_e/\partial R$ & accel., uniform plasma & accel., non-uniform plasma \\
		\hline
		9500 & 9.6e19 & 4.29 & -1.3e4 eV/m & 1.48 e18 $m^{-4}$ &  8.0e6 $m/s^2$ & 2.3e7 $m/s^2$\\
		\hline
	\end{tabular}
	\caption{Plasma and fragment state at the simulation points for rocket acceleration.}
	\label{table1}
\end{table}

\subsubsection{Two fragments separated by 2 cm and 4 cm in the transverse direction to the magnetic field and in the direction of grad B drift.}

In this section, we present numerical simulations of two fragments separated by 2 cm and 4 cm  in the transverse direction to the magnetic field and in the direction of grad B drift. For all simulations presented in this and the next sections, 1.3 mm radius fragments were ablated in a uniform plasma with $n_e = 9.6e19\, m^{-3}$ and $T_e = 9550$ eV.

\begin{figure}[h!]
	\centering
	\subfigure[2 cm transverse separation]{\includegraphics[width=.8\linewidth]{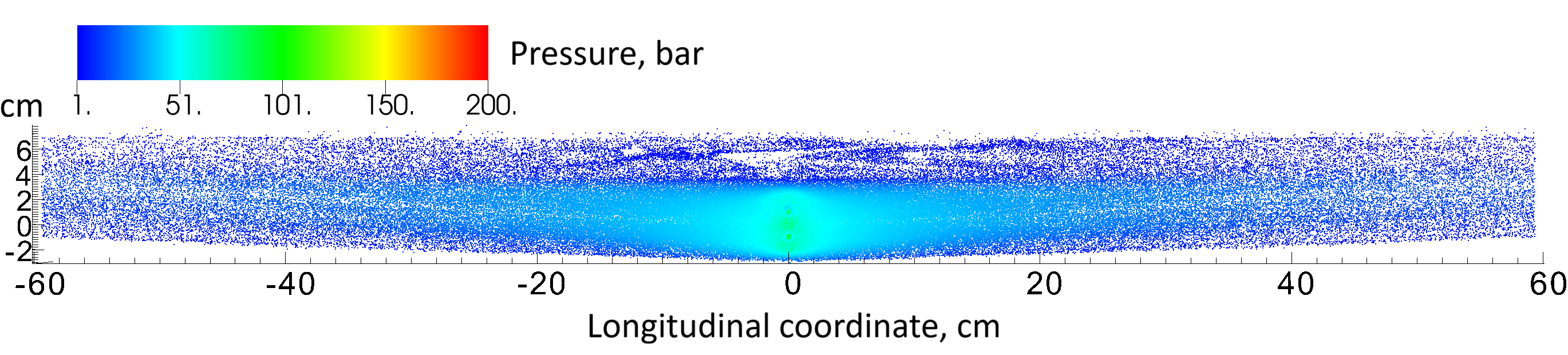}}
	\subfigure[4 cm transverse separation]{\includegraphics[width=.8\linewidth]{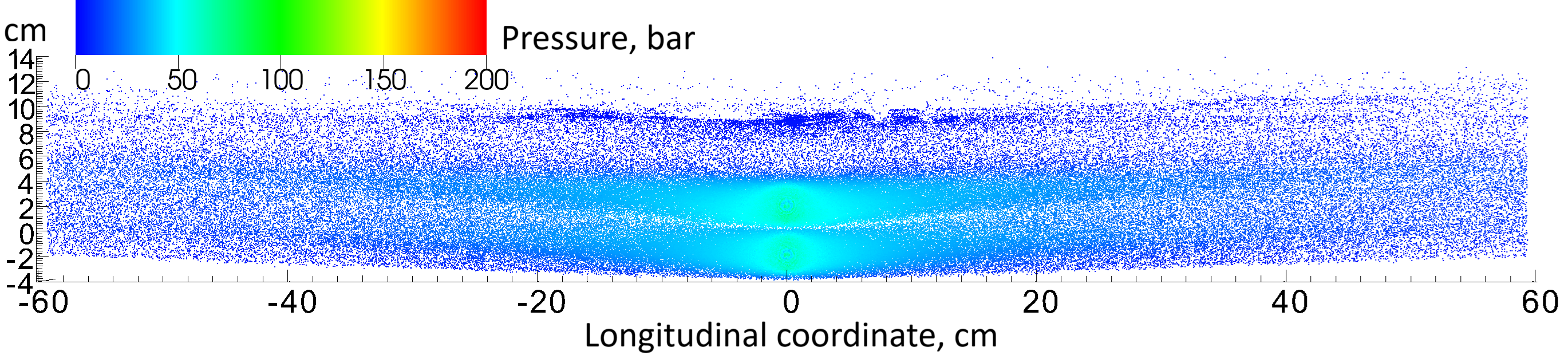}}
	\caption{Comparison of ablation processes of two fragments separated by 2 cm (a) and 4 cm (b) in the transverse direction to the magnetic field and in the direction of grad B drift. Pressure distributions (bar) are shown on thin slices of data. In all plots, $x$-axis is along the magnetic field and $y$-axis is in the direction of the major tokamak radius.}
	\label{fig:2fr_t}
\end{figure}

Figure \ref{fig:2fr_t} compares pressure distributions for the corresponding clouds.  Ablation clouds of fragments separated by 2 cm are strongly interacting with each other, resulting in the redistribution of pressure in the vicinity of each fragment. Ablation clouds look more independent if the separation distance is increased to 4 cm, but the grad-$\mathbf{B}$ drift of ablated material from the bottom fragment caused quantitative differences. In both cases, the rocket acceleration of the bottom fragment was higher compared to the acceleration of the top fragment, but the difference was larger in the case of 2 cm separation. The bottom fragment acceleration was about $8.e6 m/s^2$ and $8.3e6 m/s^2$ for the 4 cm and 2 cm separation, respectively. The corresponding numbers for the top fragments were about $6.e6 m/s^2$ and $4.e6 m/s^2$. Close fragments will be approaching each other until they reach the high pressure region in the center caused by the collision of both ablation flows.

\subsubsection{Two fragments separated by 4 cm along the same magnetic field line.}

\begin{figure}[h!]
	\centering
	\subfigure[Density, g/cc]{\includegraphics[width=.8\linewidth]{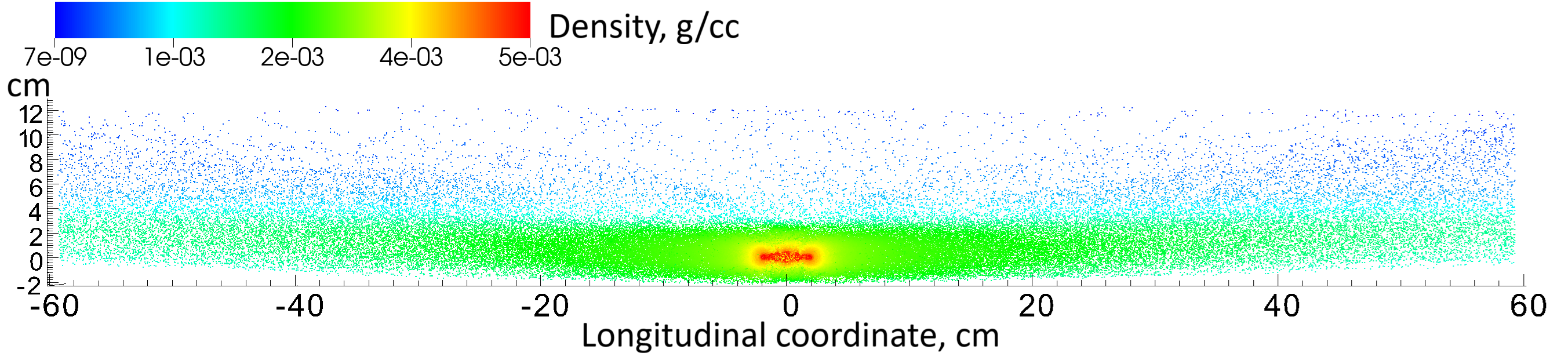}}
	\subfigure[Pressure, bar]{\includegraphics[width=.8\linewidth]{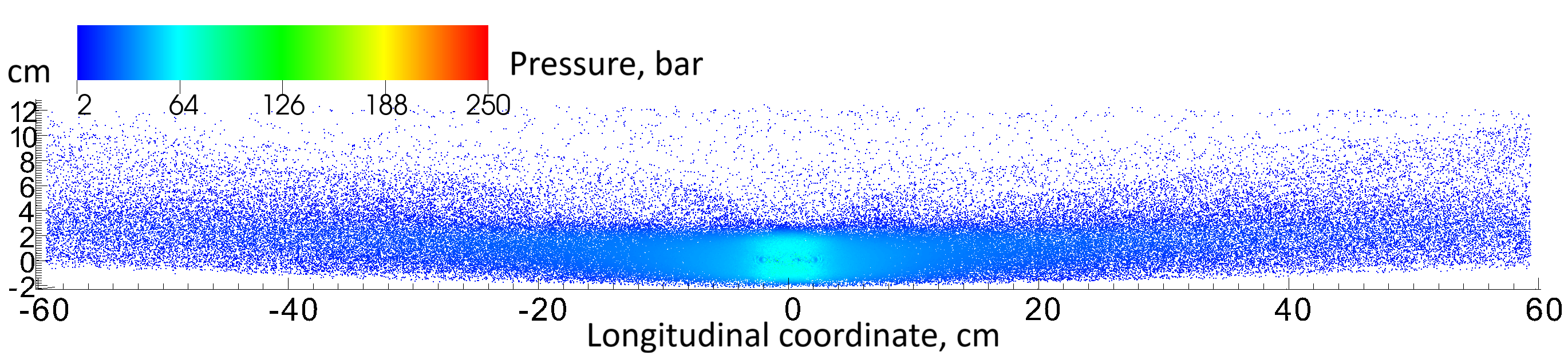}}
	\subfigure[Pressure, bar, around two ablating fragments (zoom in plot (b))]{\includegraphics[width=.8\linewidth]{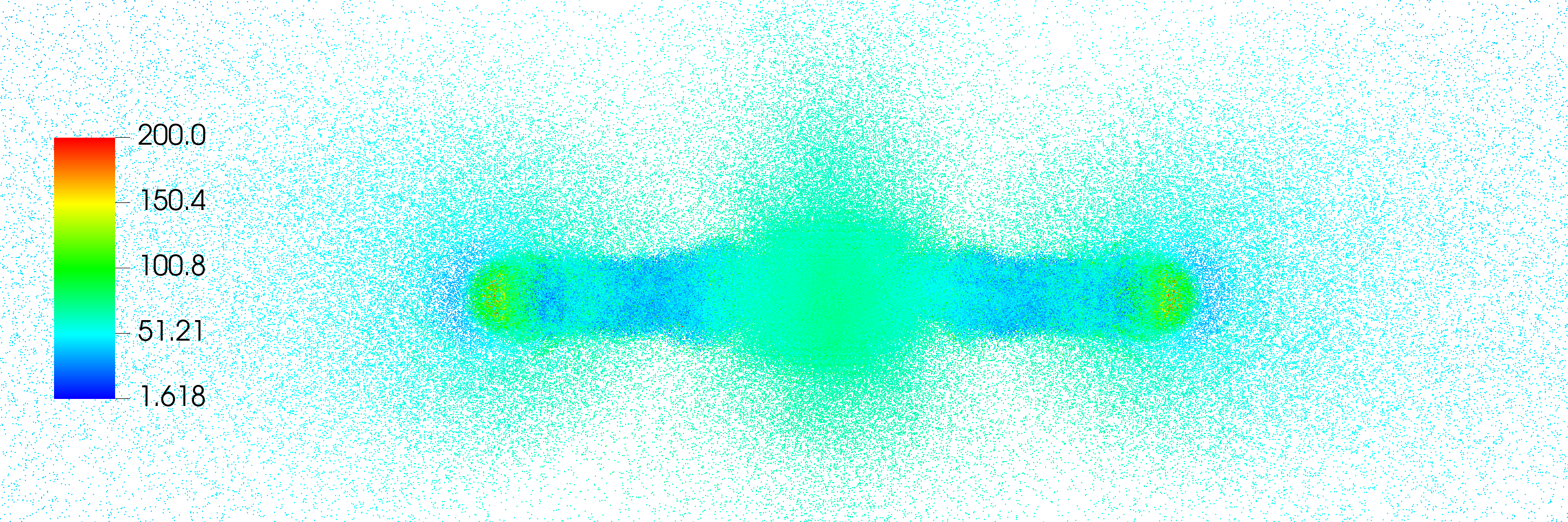}}
	\caption{Simulation of two fragments separated by 4 cm along the same magnetic field line.  In all plots, $x$-axis is along the magnetic field and $y$-axis is in the direction of the major tokamak radius.}
	\label{fig:2fr_4l}
\end{figure}

In this section, we present numerical simulations of two fragments separated by 4 cm  along the same magnetic field line crossing the pellet centers. Figure \ref{fig:2fr_4l} shows distributions of density and pressure in ablation clouds. The last image in Figure \ref{fig:2fr_4l} shows a more detailed pressure distribution around the ablating fragments. The ablated material flows from each fragment move in opposite directions and create a region of increased pressure between them.
Due to  the partial screening of the incoming electron heat flux along magnetic field lines by the neighboring fragment, the rocket acceleration significantly changes not only in the absolute value but also in its direction. In the $\nabla R$ direction, the rocket acceleration was reduced by an order of magnitude compared to the previous simulations: it was about $7.e5 m/s^2$. But the rocket acceleration of fragments towards each other was significantly higher: $7.e7 m/s^2$. Fragments will move towards each other until they reach the high pressure region caused by opposite flows. 

These observations can be qualitatively explained as follows. High pressure at the pellet fragment surface is caused by the directional heating by fast electrons streaming along magnetic field lines. The pressure difference between high- and low-field sides due to the cloud asymmetry, causing the rocket acceleration in the $\nabla R$ direction, is a lower order of magnitude effect. If one side of the heating is removed (when the two fragments on the same magnetic field line are in the each other shadow),  the pressure from outside drives the pellets towards each other. This acceleration will be reduced when the pellets enter the region of colliding ablation flows in the middle, characterized by an increased pressure.  The rocket acceleration is reduced in the $\nabla R$ direction by the redistribution of the neutral cloud density between the fragments.

However, a slight shift of one fragment away from the magnetic field line "occupied" by the other fragment by a distance larger than the fragment diameter largely eliminates this effect.  As the overall conclusion, SPI fragments will have a tendency to move towards each other. But larger separation distances between fragments in real SPI clouds and, most important, the randomness of their location make the overall effect small. 

\section{Summary and future work}

A comprehensive numerical model of the rocket acceleration of cryogenic pellets and shattered‐pellet fragments in magnetized fusion plasmas has been developed and implemented in the PELOTON Lagrangian‐particle simulation framework. The model incorporates non-uniform electrostatic charging of the ablation cloud, grad-$\mathbf{B}$ drift–induced cloud asymmetry, and realistic plasma conditions obtained from the newly developed PELOTON Cooling Module. Together, these capabilities enable quantitative prediction of pellet or fragment acceleration caused by asymmetric ablation pressures on the high- and low-field sides of the pellet.

Simulations performed for plasma parameters characteristic of JET SPI experiments qualitatively agree with  observed fragment trajectories and deviations toward the low-field side. From systematic parameter scans, a scaling law was derived that relates the pressure asymmetry—and hence rocket acceleration—to local electron temperature and density. The scaling demonstrates that rocket acceleration depends strongly on plasma temperature and more weakly on density, but is essentially independent of pellet radius within the range relevant to JET experiments.

Inclusion of 0.5 \% neon in the pellet composition was found to reduce both ablation cloud temperature, longitudinal expansion and grad-$\mathbf{B}$ drift velocities, leading to smaller, though still measurable, accelerations. We attribute the absence of rocket acceleration effect on trajectories of deuterium-neon fragments in the corresponding SPI experiments in JET to significantly changed background plasma states due to neon radiation and corresponding triggering of MHD events.

Additional numerical studies examined the influence of plasma gradients and fragment-cloud interactions. While strong local plasma gradients can amplify the rocket acceleration by factors of order 3, these effects act only over short time intervals. Simulations of multiple fragments revealed that cloud overlap in the grad-$\mathbf{B}$ direction slightly enhances the acceleration of downstream fragments, while fragments aligned along the same magnetic field line experience strong mutual attraction due to partial screening of the incoming electron flux. While the result is interesting from the ablation physics point of view, we acknowledge that the probability of two fragments remaining along the same magnetic field line during their motion is very low.

Overall, the new model captures the essential physics of pellet rocket acceleration in magnetized plasmas and extends PELOTON's capabilities. Future work will apply this framework to simulate SPI in ITER-relevant conditions, refine the inclusion of plasma diffusion in the cooling module, and establish a scaling law for rocket acceleration in ITER plasmas.

\appendix

{\cbl
\section{Simulation Points Used for Ablation Rate and Rocket Acceleration in JET}
\begin{table}[H]
\centering
\begin{tabular}{|c|c|c|c|c|c|c|}
\hline
Simulation No. & Fragment Radius (mm) & ne ($m^{-3}$)& Te (eV)& G (g/s) & $\Delta P$ (bar) & Acceleration ($ms^{-2}$)\\
\hline
 1& 0.625  &5.7e19  &554.67  &1.135  & 0.16 & 9.7e4\\
 2&0.48  &6.1e19  &786.065  &1.190  & 0.23 &1.8e5 \\
 3& 0.31 & 6.1e19 &920.823  &1.198  & 0.26 & 3.1e5\\
 4& 0.19 &6.1e19  &997.636  &.74  & 0.22 &4.3e5\\
 5&1.5  &8.5e19  &1530  &23.1  & 0.68 & 1.7e5\\
 6&1.0  &8.2e19  &1800  &19.0  & 0.70 & 2.6e5\\
 7&0.5  &7.9e19  &1982  &8.9  & 0.48 &3.6e5\\
 8&0.25  &7.8e19  &2050  &3.85  & 1.41 &2.1e6\\
 9&1.2  &8.5e19  &1230  &11.1  & 0.52 &1.6e5\\
 10&0.8  &8.5e19  &1400  &9.1  & 0.62 &2.9e5\\
 11&0.4  &8.5e19  &1550  &4.5  & 0.53 &4.9e5\\
 12&0.2  &8.5e19  &1570  &1.89  & 0.64 &1.2e6\\
 13&1.79  &8.75e19  &860.8  &9.2  & 0.39 &8.0e4\\
 14&1.47  &8.7e19  &1175.5  &13.8  & 0.55 &1.4e5\\
 15&0.4  &8.5e19  &1630.5  &5.05  & 0.55 &5.1e5\\
 16&0.23  &8.64e19  &1602.5  &2.45  & 0.80 &1.3e6\\
 17&1.0  &1.8e20  &1000  &4.04  & 0.40 &1.5e5\\
 18&1.0  &3.04e20  &465.16  &1.94 & 0.22 &8.2e4\\
 19&1.0  &2.5e20  &686.38  &3.97  & 0.43 &1.6e5\\
20&1.0  &1.26e20  &1219.33  &10.3  & 0.68 &2.5e5\\
\hline

\end{tabular}
\caption{{\cbl This table lists the simulation data points and results used to create function fits used in section \ref{sec: JET SPI Simulations}.}}
\label{table:JET_sim_data}
\end{table}
}

{\bf Acknowledgment.}
{This work has been supported in part by the ITER Organization and the ITER Disruption Mitigation Task Force via contract. The views and opinions expressed herein do not necessarily reflect those of the ITER Organization. The Swiss contribution was supported in part by the Swiss National Science Foundation.}

\bibliographystyle{unsrt} 

\end{document}